\documentclass[11pt,a4paper]{article}
\usepackage{axodraw4j}
\usepackage{graphicx}
\usepackage{jheppub}
\usepackage{subfig}
\usepackage[latin1]{inputenc}
\usepackage{amsmath}
\usepackage{amsfonts}
\usepackage{amssymb}
\usepackage{setspace}
\usepackage{mathtools}
\usepackage{pstricks}
\usepackage{color}
\providecommand{\f}[2]{\frac{{#1}}{{#2}}}
\newcommand{\da}{\ensuremath{\dot{a}}}

\newcommand{\ee}[1]{\begin{equation}#1\end{equation}}
\newcommand{\ea}[1]{\begin{align}#1\end{align}}
\newcommand{\eg}[1]{\begin{gather}#1\end{gather}}

\title{Quantum corrections to scalar field dynamics in a slow-roll space-time}

\author[a]{Matti Herranen,}
\author[b]{Tommi Markkanen,}
\author[c]{Anders Tranberg}

\affiliation[a]{Niels Bohr International Academy and Discovery Center, Niels Bohr Institute,
University of Copenhagen, Blegdamsvej 17, 2100 Copenhagen, Denmark}
\affiliation[b]{Helsinki Institute of Physics and Department of Physics, P. O. Box 64, FI-00014, University of Helsinki, Finland}
\affiliation[c]{Faculty of Science and Technology, University of Stavanger, \\4036 Stavanger, Norway}

\emailAdd{herranen@nbi.dk}
\emailAdd{tommi.markkanen@helsinki.fi}
\emailAdd{anders.tranberg@uis.no}

\abstract{We consider the dynamics of a quantum scalar field in the background of a slow-roll inflating Universe. We compute the one-loop quantum corrections to the field and Friedmann equation of motion, in both a 1PI and a 2PI expansion, to leading order in slow-roll. Generalizing the works of \cite{Sloth:2006az,Serreau:2011fu,Bilandzic:2007nb,Boyanovsky:2005sh}, we then solve these equations to compute the effect on the primordial power spectrum, for the case of a self-interacting inflaton and a self-interacting spectator field. We find that for the inflaton the corrections are negligible due to the smallness of the coupling constant despite the large IR enhancement of the loop contributions. For a curvaton scenario, on the other hand, we find tension in using the 1PI loop corrections, which may indicate that the quantum corrections could be non-perturbatively large in this case, thus requiring resummation.}


\begin{document}
\maketitle

\section{Introduction}
\label{sec:intro}

The primordial fluctuations seeding galaxy formation and observed in the CMB are expected to originate from quantum fluctuations in the energy density during inflation. In the simplest case, a single scalar field, the inflaton, dominates both the overall energy density leading to inflation and the fluctuations. A more complicated case is the curvaton scenario \cite{Linde:1996gt,Enqvist:2001zp,Lyth:2001nq,Moroi:2001ct}, where the inflaton still generates the inflationary expansion, but a second field, the curvaton, sources the fluctuations by dominating the energy density briefly during the right epoch.

Scalar fields are quantum mechanical in origin, and it is therefore important, once a given model has been established to give the right predictions for the CMB, to test the robustness of the results to the inclusion of quantum corrections, which are not necessarily negligible \cite{vanderMeulen:2007ah}. Recently this topic was revived in \cite{Weinberg:2005vy,Weinberg:2006ac} in the context of loop corrections for cosmological correlators. Quantum effects can also be examined by a (truncation of a) loop and/or gradient expansion of the effective action, and the evolution equations that derive from it. As is always the case, proper renormalization must be taken into account, and it is wise to carefully consider the vacuum state one renormalizes to. 

A large body of work has examined in particular truncations of the 1-particle-irreducible (1PI) effective action, and renormalization issues. These are typically based on a gradient expansion around the Minkowski vacuum, either through adiabatic regularisation \cite{Parker:1974qw,Ramsey:1997sa,Anderson:2005hi,MolinaParis:2000zz,Tranberg:2008ae} or at the level of the action using the Schwinger-deWitt expansion \cite{Paz:1988mt,Hu:1984js,Kirsten:1993jn,Markkanen:2012rh} for the 1PI effective action. In contrast to the 1PI expansion, in the 2-particle-irreducible (2PI) expansion one uses the dressed propagator in the Feynman diagrams. At one-loop level this corresponds to resumming an infinite series of perturbative SuperDaisy diagrams into a single mass term. The 2PI expansion in curved spacetime has been studied in \cite{Riotto:2008mv,Boyanovsky:1997cr,Ramsey:1997qc}, and for a recent more particle physics driven approach based on diagrams, see \cite{George:2012xs,George:2013iia}. 

Computation of the effective action in de Sitter space was performed in \cite{Sloth:2006nu,Sloth:2006az,Serreau:2011fu,Bilandzic:2007nb}, where in \cite{Serreau:2011fu} the 2PI-leading order truncation was used (see also \cite{Arai:2012sh,Nacir:2013xca}), which is also employed here, and in \cite{Bilandzic:2007nb} an RG improved 1PI expansion at leading order and in \cite{Boyanovsky:2005sh} 1PI to first order in slow-roll expansion was considered. 
The effective 2PI equations in dS have also recently been studied in \cite{Nacir:2014hka} with  adiabatic subtraction as the regularization method. Higher order 1PI loop effects in de Sitter space were considered in \cite{oai:arXiv.org:0904.4811,oai:arXiv.org:gr-qc/0408080,oai:arXiv.org:gr-qc/0406098, 
oai:arXiv.org:gr-qc/0204065,Onemli:2013gya}.
The main emphasis of in particular \cite{Prokopec:2002jn,Gautier:2013aoa, Garbrecht:2011gu,Garbrecht:2013coa,Sloth:2006az,Serreau:2011fu, Serreau:2013psa} was the generation of an effective mass even for massless fields, through interaction or self-interaction and non-Gaussian correlators were considered in \cite{Serreau:2013koa}. As was first shown in \cite{Burgess:2010dd,oai:arXiv.org:0912.1608} (see also \cite{oai:arXiv.org:astro-ph/9407016}), 
 in this way, the IR problems of perturbative expansion in terms of free propagators can be avoided, including certain secular time behaviour specific to FRW space-times.

In the present paper, we will compute the 1-loop quantum correction to the field and Friedmann equations of motion to leading order in slow-roll; both in the 1PI and the resummed 2PI expansions, in particular comparing when either can be trusted as an approximation. We will take the vacuum to be the slow-rolling one, and replace a gradient expansion by the slow-roll expansion. We then compute the leading quantum corrections to the slow-roll parameters and estimate their effect on the primordial power spectrum of the CMB.

After this introduction and presenting the model, we will in section \ref{sec:1PI} derive the 1PI system of equations at one-loop, computed in the slow-rolling quasi-de Sitter vacuum. In section \ref{sec:2PI1} we do the same thing for the 2PI system of equations and compare the two approaches to each other and to results in the literature. In section \ref{sec:num} we solve the equations for the leading quantum corrections to the slow-roll parameters in the case of inflaton, relating to corrections in the primordial CMB observables. We also briefly consider the role of quantum corrections in the curvaton scenario. A number of details are relegated to a set of appendices, and we conclude in section \ref{sec:conc}.

We will consider a massive $\varphi^4$ theory, with non-minimal coupling to gravity. 
Using the (+,+,+) convention of \cite{Misner:1974qy}, our $n$ space-time dimensional action has the standard matter and gravitational parts 
\ea{
S[\varphi,g^{\mu\nu}]&\equiv S_m[\varphi,g^{\mu\nu}]+S_g[g^{\mu\nu}],\nonumber\\[2mm]
\label{eq:actm}S_m[\varphi,g^{\mu\nu}]&=-\f{1}{2}\int d ^nx\sqrt{-g}~ \bigg[\partial_\mu\varphi\partial^\mu\varphi+m^2_0\varphi^2 +\xi_0 R\varphi^2+2\f{\lambda_0}{4!}\varphi^4\bigg]\\\label{eq:actg}S_g[g^{\mu\nu}]&= \int d^nx\sqrt{-g}~\bigg[\Lambda_0+\alpha_0 R+\beta_0 R^2+\epsilon_{1,0}R_{\alpha\beta}R^{\alpha\beta}+ \epsilon_{2,0}R_{\alpha\beta\gamma\delta}R^{\alpha\beta\gamma\delta}\bigg],}
where the higher order tensors in the gravitational part are required for the renormalization of the theory \cite{ParkerToms}. The subscript "0" denotes a bare quantity with the standard decomposition to a finite parameter and a counter term as, $c_0=c+\delta c$. Our working assumption will be that all the higher order gravitational terms have vanishing renormalized values, i.e. $\beta=\epsilon_1=\epsilon_2=0$, but this assumption is not needed for any of the following to go through.

Throughout this paper we will assume that our space-time has the metric $g_{\mu\nu}dx^{\mu}dx^{\nu}=-dt^2+a^2d\mathbf{x}^2$ i.e. it is of the Friedmann-Robertson-Walker (FRW) form, and therefore we will not consider the effect of the metric perturbations (neither classical nor quantized) on the dynamics of the matter fields. This choice is motivated by our desire to work with a renormalizable theory, i.e within the context of semi-classical gravity. Possibly, a more complete prescription would be to include quantized metric perturbations, however, non-perturbative resummations of infinite number of Feynman diagrams can be problematic when working with a non-renormalizable theory such as gravity. 
The loop corrections including quantized metric perturbations have been recently studied in \cite{George:2013iia} for the case of Higgs inflation, where it was found that at least for large gravity-matter coupling $\xi$ these effects may be important.

In defining the action (\ref{eq:actm}) we have chosen to neglect all operators with a mass dimensions higher than four. When assuming classical gravity this choice is problematic from an effective theory point of view, since the field values of $\varphi$ range all the way up to Planck scale and the higher order operators cannot anymore be viewed as Planck-suppressed.\footnote{Neglecting the operator $\varphi^6$ is problematic also for small-field inflationary models.} This is a manifestation of the well-known $\eta$-problem, which at the moment remains unresolved (see \cite{oai:arXiv.org:1004.3801} for a recent discussion).

\section{1PI truncation at one loop}
\label{sec:1PI}

Deriving the one-loop 1PI equations of motion for the quantized theory is a well-known procedure \cite{Peskin:1995ev}, which we perform by shifting the field operator as $\hat{\varphi}=\varphi+\hat{\phi}$, with the expectation value written as $\langle\hat{\varphi}\rangle\equiv\varphi$, and expand (\ref{eq:actm}) around $\hat{\phi}=0$ giving to quadratic order\footnote{As explained in \cite{Markkanen:2013nwa}, since the counter terms are already of one-loop order their inclusion in the quantum correction is a two-loop effect and hence beyond the one-loop approximation.}
\ea{\label{eq:expS}S_m[\hat{\varphi},g^{\mu\nu}]=&-\f{1}{2}\int d ^nx\sqrt{-g}~ \bigg[\partial_\mu\varphi\partial^\mu\varphi+m^2_0\varphi^2 +\xi_0 R\varphi^2+2\f{\lambda_0}{4!}\varphi^4\bigg]\nonumber \\ &-\f{1}{2}\int d ^nx\sqrt{-g}~\hat{\phi}\bigg[-\Box +m^2+\xi R+\f{\lambda}{2}\varphi^2\bigg]\hat{\phi}+\cdots.}
From now on for simplicity we will define the one-loop effective mass as
\ee{M^2 \equiv m^2+\xi R+\f{\lambda}{2}\varphi^2,}
with which the equation of motion for the fluctuation operator $\hat{\phi}$ is simply
\ea{\bigg[-\Box +M^2\bigg]\hat{\phi}=0,\label{eq:eomR3}}
where $\Box \equiv 1/\sqrt{-g}\,\partial_\mu (\sqrt{-g}\,\partial^\mu)$. Assuming the FRW metric allows us to write equation (\ref{eq:eomR3}) in a more familiar way by using the properly normalized ansatz 
\eg{\hat{\phi}=\int d^{n-1}k\big [{a}_\mathbf{k} u_\mathbf{k}+{a}^*_\mathbf{k} u_\mathbf{k}^*\big],\qquad u_\mathbf{k}({\bf x},t)=\f{1}{\sqrt{2(2\pi)^{n-1}a^{n-1}}}h_\mathbf{k}(t)e^{i\mathbf{k}\cdot \mathbf{x}},\label{eq:fielddef}}
with the standard commutation relations for the operators
\ee{[\hat{a}_\mathbf{k},\hat{a}_{\mathbf{k}'}]=[{\hat{a}}^\dagger_\mathbf{k},{\hat{a}}^\dagger_{\mathbf{k}'}]=0,\qquad[{\hat{a}}_\mathbf{k},{\hat{a}}^\dagger_{\mathbf{k}'}]=\delta^{n-1}(\mathbf{k}-\mathbf{k}'),\label{eq_com}}
for which (\ref{eq:eomR3}) becomes
\eg{ \ddot{h}_\mathbf{k}(t)+\bigg[-\bigg(\f{n-1}{2}\bigg)^2H^2-\f{n-1}{2}\dot{H}+\f{\mathbf{k}^2}{a^{2}}+M^2\bigg]h_\mathbf{k}(t)=0\label{eq:mode}
,}
where $H\equiv \da/a$ is the Hubble constant.

\subsection{Momentum modes to first order in slow-roll}
\label{sec:1pisol}

Next we will assume that the classical metric background is slow-rolling, in the sense that the deviation from pure exponential expansion (de Sitter space), can be written as an expansion in the small quantities $\epsilon$ and $\delta_H$,\footnote{We use the subscript $H$ to distinguish $\delta_H$ from $\delta$, defined below in (\ref{eq:defdel}).}
\ee{\epsilon\equiv-\f{\dot{H}}{H^2}\label{eq:defeps}\,,\qquad\qquad \delta_H=\f{\ddot{H}}{2H\dot{H}}\,.}
From the definition (\ref{eq:defeps}) one finds for the time-derivative of $\epsilon$
\ee{\dot{\epsilon}=2\epsilon\big(\epsilon+\delta_H\big)H\,.\label{dot_eps}}
such that $\epsilon$ is approximately constant on the time scale $1/H$. We assume that the same is true for $\delta_H$ as well. Then, by using the definitions 
\eg{{h}_{\mathbf{k}}\equiv \sqrt{\f{\pi}{2H(1-\epsilon)}}\bar{h}_{\mathbf{k}},\qquad\quad x \equiv \f{\vert\mathbf{k}\vert}{a H(1-\epsilon)},}
Eq.~(\ref{eq:mode}) can be written to quadratic order in slow-roll parameters $\epsilon$ and $\delta_H$ as 
\ee{{x}^2\f{d^2 \bar{h}_{\mathbf{k}}(t)}{d{x}^2}+{x}\f{d \bar{h}_{\mathbf{k}}(t)}{d{x}}+\big({x}^2-{\nu}^2\big)\bar{h}_{\mathbf{k}}(t)=0\,,\label{eq:bessel}}
where
\ee{{\nu}^2 \equiv
\frac{(n-1)^2}{4} + \frac{(n-1)(n-2)}{2}\epsilon + \frac{3n^2-10n+4}{4}\epsilon^2
- \f{M^2}{H^2}(1+2\epsilon+3\epsilon^2) - \delta_H \epsilon \,.\label{eq:def}
}
In the limit $\epsilon\rightarrow 0$ and constant $\nu$ this equation is the standard Bessel equation which has the Bunch-Davies \cite{Bunch:1978yq} vacuum solution\footnote{This equation is often written in terms of conformal time $ dt=a\,d\eta$: \newline $f''_{\mathbf{k}}(\eta)+\big[k^2+(\nu^2-1/4)/\eta^2\big]f_{\mathbf{k}}(\eta)=0$, with $u_\mathbf{k} = a^{\f{n-2}{2}}f_{\mathbf{k}}(\eta)$.} ${h}_{\mathbf{k}}(t)=\sqrt{\pi/(2H)}\,H^{(1)}_{{\nu}}({x})$, where $ H^{(1)}_{{\nu}}$ is the Hankel function of the first kind.
For the boundary conditions for the mode functions we impose that the mode corresponds to the positive frequency mode at high momentum i.e
\ee{{h}_{\mathbf{k}}(t)\rightarrow \f{e^{-i\int^t \omega(t') dt'}}{\sqrt{\omega(t)}},\qquad \omega(t)\rightarrow \f{k}{a},}
at $k\rightarrow \infty$ where $k \equiv |\mathbf{k}|$.
Using the asymptotics of the Hankel function and the above boundary conditions we get the approximate
 solution (see also \cite{Weinberg:2008zzc})
\ee{h_{\mathbf{k}}(t)=\sqrt{\f{\pi}{2H(1-\epsilon)}}\bigg[C_1(k) H^{(1)}_{{\nu}}({x})+ C_2(k)H^{(2)}_{{\nu}}({x})\bigg]\label{eq:sol},}
with $C_{1,2}$ having the property $C_1(k) \rightarrow 1$ and $C_2(k) \rightarrow 0$ when $k\rightarrow \infty$. For simplicity, we will here make the choice
\ee{C_1(k) \equiv 1\,,\qquad\qquad C_2(k) \equiv 0\,,}
which obviously reduces to de Sitter symmetric Bunch-Davies vacuum solution in the limit $\epsilon \to 0$. The solution (\ref{eq:sol}) satisfies equation (\ref{eq:bessel}) up to terms proportional to time-derivative of the index,
 $\dot\nu \sim {\cal O}(\epsilon^2, \epsilon\delta_H, \epsilon\delta )$,\footnote{We have also assumed that $\dot\delta \sim {\cal O}(\epsilon^2, \epsilon\delta_H, \epsilon\delta )$.} where
\ee{\delta\equiv \f{M^2}{H^2}\label{eq:defdel}.} 
Hence, the neglected terms are indeed subleading assuming that $\delta \ll 1$. As we will later see, for a typical single-field inflaton scenario $\delta \sim \delta_H \sim \epsilon$. 

\subsection{1PI effective equations of motion}
\label{sec:effeq}

The renormalized quantum corrected equations of motion can be derived to one-loop order in 1PI expansion without explicit reference to the effective action \cite{Markkanen:2013nwa}. The advantage of this approach is that the finite parts of the counter terms will be suited for the particular space-time geometry of interest.

The effective equations of motion, i.e. the field equation and the Einstein equation result from the variations of the action
\ee{\label{eq:var2}\bigg\langle\f{\delta S[\hat{\varphi},g^{\mu\nu}]}{\delta\hat{\varphi}(x)}\bigg\rangle=0
\qquad{\rm and}\qquad\bigg\langle\f{\delta S[\hat{\varphi},g^{\mu\nu}]}{\delta g^{\mu\nu}(x)}\bigg\rangle=0,}
respectively. For the action in (\ref{eq:actm}) the field equation (\ref{eq:var2}) becomes 
\ee{\bigg[-\Box +m^2+\delta m^2+(\xi +\delta\xi )R\bigg]\varphi+\f{\lambda+\delta\lambda}{3!}\varphi^3+\f{\lambda}{2}\varphi\langle\hat{\phi}^2\rangle=0.\label{eq:eom1}}
Similarly, we can write for the Einstein equation
\ee{\f{1}{8\pi G}(\Lambda g_{\mu\nu}+ G_{\mu\nu}) = -\f{2}{\sqrt{-g}}\bigg\langle\f{\delta}{\delta g^{\mu\nu}} \big(S_m[\hat{\varphi},g^{\mu\nu}]+S_{\delta g}[g^{\mu\nu}]\big)\bigg\rangle \equiv T_{\mu\nu} \label{eq:nonrE},}
where we have set $\Lambda\rightarrow -\Lambda/(8\pi G)$ and $\alpha \rightarrow 1/(16\pi G)$ in order to match with standard convention, and we further split the energy-momentum tensor into classical, quantum and counter-term contributions, respectively:
\ea{T_{\mu\nu} &\equiv T_{\mu\nu}^C+\langle \hat{T}^Q_{\mu\nu}\rangle+\delta T_{\mu\nu}
\nonumber\\
&\equiv T_{\mu\nu}^C+\langle\underline{\hat{T}^{Q}_{\mu\nu}}\rangle,}
with
\ea{T_{\mu\nu}^C&=-\f{g_{\mu\nu}}{2}\bigg[\partial_\rho\varphi\partial^\rho\varphi+m^2\varphi^2+2\f{\lambda}{4!}\varphi^4 \bigg]+\partial_\mu\varphi\partial_\nu\varphi\nonumber \\&+\xi\big[G_{\mu\nu}-\nabla_\mu\nabla_\nu+g_{\mu\nu}\Box\big]\varphi^2,\label{eq:clasEM}
\\[5mm]
\langle \hat{T}_{\mu\nu}^Q\rangle&=-\f{g_{\mu\nu}}{2}\bigg[\f{\partial}{\partial x_\rho}\f{\partial}{\partial y^\rho} +M^2 \bigg]G(x,y)\big\vert_{x=y}+\f{\partial}{\partial x^\mu}\f{\partial}{\partial y^\nu}G(x,y)\big\vert_{x=y}\nonumber \\&+\xi\big[R_{\mu\nu}-\nabla_\mu\nabla_\nu+g_{\mu\nu}\Box\big]G(x,x)\label{eq:quantumEM},}
and
\ea{\delta T_{\mu\nu}&\equiv\delta T^m_{\mu\nu}+\delta T^g_{\mu\nu},\nonumber \\[3mm]
\delta T^m_{\mu\nu}&=-\f{g_{\mu\nu}}{2}\bigg[
\delta m^2\varphi^2+2\f{\delta\lambda}{4!}\varphi^4 \bigg]+\delta\xi\big[G_{\mu\nu}-\nabla_\mu\nabla_\nu+g_{\mu\nu}\Box\big]\varphi^2,\label{eq:cters1} \\
\delta T^g_{\mu\nu}&=-\f{2}{\sqrt{-g}}\f{\delta S_{\delta g}[g^{\mu\nu}]}{\delta g^{\mu\nu}}=g_{\mu\nu}\delta\Lambda-2\delta\alpha G_{\mu\nu} -2\delta\beta~^{(1)}H_{\mu\nu}-2\delta\epsilon_{1}~^{(2)}H_{\mu\nu}-2\delta\epsilon_{2}H_{\mu\nu},\label{eq:cters2}}
with the propagator defined as $G(x,y)=\langle 0\vert \hat{T}\big\{\hat{\phi}(x)\hat{\phi}(y)\big\}\vert 0\rangle$ where $\hat{T}$ denotes time-ordering.\footnote{We note that although we consider an out-of-equilibrium setup, as long as we compute the local correlator, truncating at one loop, we do not need to worry about the Schwinger-Keldysh contour, and distinguishing between field variables living on the upper and lower branch. Our $G(x,x)$ is $G^{++}(x,x)$ in the notation of \cite{Riotto:2008mv}, and the statistical propagator $F(x,x)$ in the notation of \cite{Tranberg:2008ae} and related.} Higher order gravitational tensors in $\delta T^g_{\mu\nu}$ result from the variation of the gravitational counter term $S_{\delta g}[g^{\mu\nu}]$ and their expressions in a FRW space-time can be found in appendix \ref{sec:appA}. 

The next step is to obtain the expressions for the variance $\langle\hat{\phi}^2\rangle=G(x,x)$ and the quantum energy-momentum tensor $\langle \hat{T}^Q_{\mu\nu}\rangle$. Our calculation of the loop integrals follows closely the steps outlined in \cite{Boyanovsky:2005sh,Serreau:2011fu} and here we merely sketch the derivation leaving the details to appendices \ref{sec:quant},\ref{sec:loopEM} and \ref{sec:line}. Our analysis is be based on an expansion in the small parameters $\epsilon$, $\delta_H$ and $\delta$,\footnote{To leading order our $\delta$ is proportional, but not identical, to the second potential slow-roll parameter $\delta_V=M_{\rm pl}^2\f{V''}{V}= \f{M^2}{3H^2}$. In section \ref{sec:2PI1} this connection is less trivial since there our definition of delta comes via the re-summed effective mass.} 

The procedure consists of first writing the momentum integrals with the variable 
$x=\vert\mathbf{k}\vert / (a H(1-\epsilon))$, 
and splitting the integration into three regions 
\ee{x<\kappa_{\rm IR},\qquad \kappa_{\rm IR}<x<\kappa_{\rm UV},\qquad \kappa_{\rm UV}<x,}
with the parameters 
\ee{\kappa_{\rm IR}\ll1\ll\kappa_{\rm UV}.} 
Contrary to \cite{Boyanovsky:2005sh,Serreau:2011fu}, for the ultraviolet contribution we use dimensional regularization instead of a cut-off, which would introduce divergences that cannot be removed by covariant counter terms \cite{Hollenstein:2011cz} (and references therein). The momentum splitting procedure also has the desirable feature that the infrared region is identical in both regularization methods.

From the formula (\ref{eq:fullooppi}) in Appendix \ref{sec:loopEM} we find the result for the equal-time correlator
\ee{G(x,x) \equiv \langle\hat{\phi}^2\rangle=\f{H^2}{8\pi^2}\bigg\{(-\delta-\epsilon+2)\bigg[\f{1}{4-n}-\log\left(\frac{H}{\mu'}\right)\bigg]+\frac{3}{\delta-3 \epsilon +3 \epsilon ^2+\delta_H \epsilon  }
\bigg\}
\label{eq:full_loop},}
where 
$\mu'$ is an arbitrary renormalization scale and according to our approximation we have included the leading infrared terms and neglected the linear orders in $\epsilon$, $\delta_H$ and $\delta$, except when appearing with the logarithm, as explained in section \ref{sec:quant}.
 
Similarly, the result for the quantum energy-momentum from (\ref{eq:energydensity2}) is
\ea{\langle\hat{T}^Q_{\mu\nu} \rangle&=-g_{\mu\nu}\frac{H^4}{32 \pi ^2 }\left\{\left(-\delta^2-4\delta\epsilon+2 \delta +6 \epsilon \right)\left[\frac{1}{4-n}-\log\left(\frac{H}{\mu }\right)\right]+\frac{6\delta}{\delta-3 \epsilon +3 \epsilon ^2+\delta_H \epsilon  } \right\}\nonumber \\&+\xi\big[R_{\mu\nu}-\nabla_\mu\nabla_\nu+g_{\mu\nu}\Box\big]\langle\hat{\phi}^2\rangle,
\label{eq:energydensity}}
with $\mu = \mu'\,\text{exp}\big[\big(-1 + 2\gamma_e  - 2\log(\pi)\big)/4\big]$.
Note that in (\ref{eq:energydensity2}) we have explicitly written all the contributions in terms of $\delta$'s and $\epsilon$'s.

\subsection{Cancellation of divergences}
\label{sec:removal}

In this section we will not be interested in the finite parts of the renormalization constants, which will be fixed by specifying the renormalization conditions later on in section \ref{sec:conditions}.

In order to have consistent results, the cancellation of the divergent $1/(4-n)$ poles in the results (\ref{eq:full_loop}) and (\ref{eq:energydensity}) must be achieved via the counter terms in (\ref{eq:eom1}), (\ref{eq:cters1}) and (\ref{eq:cters2}). 
It is a straightforward calculation to derive the divergent counter terms, and they are listed in Appendix \ref{Sec:C-terms}. By defining
\ee{\theta \equiv \delta-3 \epsilon +3 \epsilon ^2+\delta_H \epsilon,
\label{theta_def}} 
the finite scalar field equation of motion (\ref{eq:eom1}) is then given by
\ea{\ddot{\varphi}&+3H\dot{\varphi}+\xi R\varphi+m^2\varphi+\f{\lambda}{6}\varphi^3\nonumber \\&+\f{\lambda\varphi H^2}{16\pi^2}\bigg\{(\delta+\epsilon-2)\log\left(\frac{H}{\mu}\right)+\frac{3}{\theta}
\bigg\}=0.\label{eq:eomfin1}}
Similarly, the finite quantum energy-momentum tensor reads
\ea{\langle\underline{\hat{T}^Q_{00}}\rangle =-a^2\langle\underline{\hat{T}^{Q}_{ii}}\rangle_{\text{1-loop}} =\frac{H^4}{32 \pi ^2 }
\bigg\{
&\Big(\delta ^2-4\delta\epsilon-2 \delta -6 \epsilon +12\xi(2-\delta+\epsilon-\delta\epsilon) \Big)
\log\left(\frac{H}{\mu }\right)
\nonumber\\
+ &~6\frac{ \delta -6 \xi}{\theta}
\bigg\},
\label{eq:finEM}}
The Einstein equation (\ref{eq:nonrE}) can then be written as two Friedmann equations\footnote{Here we have used the reduced mass defined as $8\pi G\equiv1/M_{\rm pl}^2$}
\ea{3H^2& = \frac{1}{M_{\rm pl}^2}\left[
T_{00}^C+\langle\underline{\hat{T}^{Q}_{00}}\rangle_{\text{1-loop}}
\right]\label{eq:Fried_fin1}
\\a^2\big(-3H^2+2\epsilon H^2\big)& = \frac{1}{M_{\rm pl}^2}\left[
{T_{ii}^C}+\langle\underline{\hat{T}^{Q}_{ii}}\rangle_{\text{1-loop}}
\right],\label{eq:Fried_fin2}}
where the classical energy-momentum tensor $T_{\mu\nu}^C$ was defined in (\ref{eq:clasEM}) and the underline signifies a finite contribution with the counter terms included.

Considering first the scalar field equation (\ref{eq:eomfin1}), we notice that because $\delta=M^2/H^2$, with $M^2=M^2(\varphi)$, the quantum corrections amount to a complicated effective potential, but no corrections to the kinetic term (at this order in slow-roll). They are all proportional to $\lambda$, and for the case when $\varphi$ is the inflaton, $\lambda$ is typically very small and the correction may be negligible. When $\varphi$ is a spectator field, $\lambda$ is unconstrained, and the corrections can be large. 

We also see that the denominator $\theta$ arising from the IR part of the loop integral is always small (because $\epsilon$, $\delta_H$ and $\delta$ are assumed to be small) and therefore the quantum correction gets enhanced. This IR enhancement is even stronger when $\delta \approx 3\epsilon$, which is the case for instance for a massive $\varphi^4$ inflaton with the mass term dominating the potential, as we shall see in section \ref{sec:num}. 

Next, considering the Friedmann equation(s), we see that there is a quantum contribution to the energy density and the pressure, which looks like a potential term. It is a function of the field $\varphi$ through $\delta$, and a function of the instantaneous $\epsilon$ and $H$. The same correction enters in the energy density and the pressure, and does not involve the kinetic or gradient terms for the scalar field. In \cite{Markkanen:2012rh,Markkanen:2013nwa}, it was found that expanding up to four gradients around Minkowski space, the corrections to the Friedmann equations involve kinetic terms (derivatives of $\varphi$), and the new contributions in energy density and pressure are no longer the same. This for instance prevented manipulations similar to the classical slow-roll equations to go through. 

Also in the present case (which amounts to including only two gradients), we see that the quantum correction to the potential force in the field equation does not follow from simple variation from the quantum contribution in the Friedmann equation. Again, this prevents us from using a full analogy with the slow-roll formalism.

\subsection{Renormalization conditions}
\label{sec:conditions}

In this section we will impose the renormalization conditions fixing the finite parts of the counter terms\footnote{Technically, a generic bare constant $c_0$ can be split into a finite part and a divergent counter term as $c_0=c+\delta c$, and subsequently the finite part $c$ can be split into a physical constant and a finite counter term as $c=c_{\textbf{ph}}+\tilde{\delta}c$. However, it is important to notice that in the 1-loop approximation the counter terms only enter through the constants in the classical contributions to the equations of motion.} by matching the effective potential (or rather the field equations of motion) to a classical potential at a specific renormalization point, denoted by
\ee{\mu_0 = (\varphi_0,\;H_0,\;\epsilon_0,\;\dot{\varphi}_0,\;\ddot{\varphi}_0). \label{eq:incond}}
The quantities in (\ref{eq:incond}) must form a solution to the equations of motion, and hence they are not completely independent. A natural choice in accordance with the slow-roll approximation would be to assume that the field is falling at approximately terminal velocity at the renormalization point, i.e. to set 
\ee{\ddot{\varphi}_0=0
, \label{eq:incond2}}
which allows one to solve $\dot{\varphi}_0$ form the field equation of motion (\ref{eq:eomfin1}). Furthermore, one could use the Friedmann equations to solve $H_0$ and $\epsilon_0$, so that eventually all quantities of interest could be expressed in terms of just $\varphi_0$. However, we refrain from making any such choices for the time being.

To begin, we write the field equation of motion (\ref{eq:eomfin1}) symbolically as
\ee{\ddot{\varphi}+3H\dot{\varphi}+\f{\partial V(\varphi,H,\epsilon)}{\partial \varphi}=0,
\label{scalar_eq_gen}} 
where the potential $V(\varphi,H,\epsilon)$ is split into a physical part and finite counter terms:
\ee{
V(\varphi,H,\epsilon)\equiv V(\varphi,H,\epsilon)_{\textbf{ph}}+\tilde{\delta} V(\varphi,H,\epsilon),}
with
\ee{\tilde{\delta} V(\varphi,H,\epsilon)=\tilde{\delta} \sigma\varphi+\f{\tilde{\delta} m^2}{2}\varphi^2+\f{\tilde{\delta}\xi}{2}R\varphi^2+\f{\tilde{\delta} \eta}{3!}\varphi^3+\f{\tilde{\delta} \lambda}{4!}\varphi^4.}
For completeness, we have introduced counter terms for one- and three-point couplings, even though these terms are not present classically and they are not needed for removing the quantum divergences.

The renormalization method we will use was explained in detail in \cite{Markkanen:2013nwa}, however, here the quantity of interest is the scalar field potential $V(\varphi,H,\epsilon)$ instead of the energy-density. Our prescription for the finite parts of the counter terms will be to renormalize $V(\varphi,H,\epsilon)$ to match the classical potential
\ee{V_C(\varphi) = \f{1}{2}m_{\textbf{ph}}^2\varphi^2+\f{\lambda_{\textbf{ph}}}{4!}\varphi^4
 \label{class_potential}}
at the renormalization point $\mu_0$, expressed by the conditions\footnote{Note that the actual effective potential $V(\varphi,H,\epsilon)$ need not be computed, since only its $\varphi$-derivative 
 appears in the renormalization conditions (\ref{eq:finrenorm}).}
\ea{\f{\partial V(\varphi,H,\epsilon)}{\partial{\varphi}} \bigg\vert_{\mu_0}&=m^2_{\textbf{ph}} {\varphi_0 }+\frac{\lambda_{\textbf{ph}}  {\varphi_0 }^3}{6},&\f{\partial^2  V(\varphi,H,\epsilon)}{\partial\varphi^2}\bigg\vert_{\mu_0}&=m^2_{\textbf{ph}} +\frac{\lambda_{\textbf{ph}}  {\varphi_0 }^2}{2}, \nonumber \\\f{\partial^3V(\varphi,H,\epsilon)}{\partial\varphi^3}\bigg\vert_{\mu_0}&=\lambda_{\textbf{ph}}\varphi_0,&\f{\partial^4V(\varphi,H,\epsilon)}{\partial\varphi^4}\bigg\vert_{\mu_0}&=\lambda_{\textbf{ph}},\nonumber \\\f{\partial^4 V(\varphi,H,\epsilon)}{\partial H^2\partial \varphi^2}\bigg\vert_{\mu_0}&=0.\label{eq:finrenorm}}
With this procedure we can solve for the finite parts of the counter terms to get the renormalized equation of motion
\ea{\ddot{\varphi}+3H\dot{\varphi}&+ \Delta \sigma+(m_\mathbf{ph}^2+\Delta m^2)\varphi+\Delta\xi R\varphi+\f{1}{2}\Delta\eta\varphi^2+\f{\lambda_\mathbf{ph}+\Delta\lambda}{6}\varphi^3\nonumber \\&+\f{\lambda_\mathbf{ph}\varphi H^2}{16\pi^2}\bigg\{(\delta_{\mathbf{ph}}+\epsilon-2)\log\left(\frac{H}{H_0}\right)+\frac{3}{\theta_{\mathbf{ph}}}
\bigg\}=0,\label{scalar_eq_fin}}
where $\delta_\mathbf{ph}$ denotes $\delta$ with all the constants replaced by the physical ones: $m^2 \to m^2_\mathbf{ph}$, etc. and similarly for $\theta_{\mathbf{ph}}$. The quantum induced $\Delta$-terms are finite constants depending on the physical parameters
 $m^2_\mathbf{ph}$ and $\lambda_\mathbf{ph}$ and the renormalization point $\mu_0$.
The explicit expressions for the $\Delta$'s assuming terminal velocity condition (\ref{eq:incond2}) can be found in appendix \ref{Sec:C-terms}.

Next, we consider the Friedman equations (\ref{eq:Fried_fin1}-\ref{eq:Fried_fin2}), which upon including finite counter terms read
\ea{3H^2& = \frac{1}{M_{\rm pl}^2}\left[
T_{00}^C+\langle\underline{\hat{T}^{Q}_{00}}\rangle_{\text{1-loop}}
+\tilde{\delta}T_{00}\right],\label{eq:Friedd1}\\
a^2\big(-3H^2+2\epsilon H^2\big)& = \frac{1}{M_{\rm pl}^2}\left[
{T_{ii}^C}+\langle\underline{\hat{T}^{Q}_{ii}}\rangle_{\text{1-loop}}+\tilde{\delta}T_{ii}
\right].\label{eq:Fried2}}
We choose to renormalize the cosmological constant such that at the renormalization point $\varphi=\varphi_0$ the energy density coincides with the classical result:
\ee{T_{00}\big\vert_{\mu_0}=T_{00}^C\big\vert_{\mu_0}=\f{1}{2}\dot{\varphi}^2_0 + V_C(\varphi_0).
\label{Lambda_finrenorm}}
By using Eq.~(\ref{eq:finEM}) we then get from equations (\ref{eq:Friedd1}-\ref{eq:Fried2}) a boundary condition and a dynamical equation, respectively, for the case of the minimal coupling $\xi_{\textbf{ph}}=0$:\footnote{We note that the counter terms are proportional to loop contributions and hence, according to our approximation, we must neglect the time-derivatives of $\delta = M^2/H^2$ in these contributions. The finite counter-term of the Einstein tensor, $\tilde{\delta}\alpha$, can be set to zero, while the finite counter terms of higher order gravity operators are negligible.}
\ea{3H_0^2& =\f{1}{M_{\rm pl}^2}\bigg(\f{1}{2}\dot{\varphi}^2_0 + V_C(\varphi_0)\bigg),\label{eq:inE}\\
\epsilon H^2&=\f{\dot{\varphi}^2}{2 M_{\rm pl}^2}.\label{eq:newfried}}
So eventually we have recovered the classical relation (\ref{eq:newfried}) connecting the field derivative $\dot\varphi$ to the slow-roll parameter $\epsilon$. On the other hand, the Friedmann equation (\ref{eq:Friedd1}) involves quantum corrections as  the energy density off the renormalization point $\mu_0$ is given by\footnote{Here we have neglected the subleading logarithmic quantum corrections for brevity.}  
\ee{
T_{00} = T_{00}^C + T_{00}^Q - T_{00}^Q\Big|_{\mu_0},
\label{full_energy}
}
with
\ee{
T_{00}^Q \approx \Delta V(\varphi) + \frac{3 H^4 \delta_\mathbf{ph} }{16 \pi ^2 \theta_\mathbf{ph}}\,,
\label{full_energyQ}
}
where we have defined
\ee{\Delta V(\varphi) \equiv \Delta \sigma\varphi + \frac{1}{2}\Delta m^2 \varphi^2 + 3\Delta\xi H^2\varphi^2 + \f{\Delta\eta}{3!}\varphi^3 + \frac{\Delta \lambda}{4!}\varphi^4\,.
\label{DelV}}
The dominant IR parts of the quantum corrections in the results (\ref{scalar_eq_fin}-\ref{DelV}) are in agreement with \cite{Boyanovsky:2005sh}.

To give a rough estimate for the size of quantum corrections for $\varphi^4$-theory of inflation, we choose as an example $m_\mathbf{ph}^2 = 0$ and $\varphi_0=22 M_{\rm pl}$, such that the renormalization point corresponds to approximately $60$ e-foldings before the end of inflation in the standard $\varphi^4$-theory of inflation ($\varphi_0^2 \approx 8(N+1)M_{\rm pl}^2$). In this case we get $\Delta\lambda/\lambda_\mathbf{ph} \sim 10^3 \lambda_\mathbf{ph}$,
which is very small for the physically viable value $\lambda_{\mathbf{ph}}\sim 10^{-12}$. 
The other $\Delta$'s as well as the quantum terms in the second row of the field equation (\ref{scalar_eq_fin}) give similar size corrections indicating that the quantum corrections may be ignored to a good approximation for the $\varphi^4$-theory of inflation. Similarly, in the massive case with $m_\mathbf{ph}^2 \sim 10^{-11} M_{\rm pl}^2$ and $\lambda_{\mathbf{ph}}\sim 10^{-15}$ with $\varphi_0=16 M_{\rm pl}$, again corresponding to roughly $60$ e-foldings before the end of inflation, we get $\Delta m^2 / m_\mathbf{ph}^2 \sim 10^{-11}$ and similar magnitudes for the other quantum corrections. Below in section \ref{sec:num}, we will estimate the size of the quantum corrections to the slow-roll parameters $\epsilon$ and $\delta_H$, which contribute directly to the observables of the primordial power spectrum of the CMB. 

More generally, we observe that the quantum corrections in the equation of motion (\ref{scalar_eq_fin}) remain perturbatively small provided that 
\ee{
\theta_{\mathbf{ph}} \gg \frac{\sqrt{\lambda_{\mathbf{ph}}}}{4\pi}\,,
\label{1PI_limit}}
which can be seen as the validity criterion for the 1PI effective action. Indeed, in the regime 
$\theta \lesssim \sqrt{\lambda_\mathbf{ph}}/(4\pi)$
it is expected that SuperDaisy resummation of the self-mass $M$ becomes important \cite{Serreau:2011fu}.
 This effect can be accommodated at one-loop level by the 2PI Hartree truncation, which we will consider next. 

\section{2PI truncation at one loop}
\label{sec:2PI1}

One may encounter infrared divergencies in perturbation theory, as a result of writing the expansion in terms of a free propagator, with small or zero mass. Although in the exact theory, a dynamical mass is generated to remove such divergences, at a finite order in a perturbative expansion, they may appear and render the results unreliable. This does not mean that infrared physics is irrelevant, and in the exact theory, what look alike divergences may in fact add up to interesting and crucial physical effects.

\begin{figure}[h]
\begin{center}
\fcolorbox{white}{white}{
  \begin{picture}(186,66) (0,-31)
    \SetWidth{1.3}
    \SetColor{Black}
    \Arc(22,18)(16,90,450)
    \Arc(22,-14)(16,180,540)
    \Line[dash,dashsize=5](116,-30)(180,-30)
    \Arc(148,-14)(16,90,450)
  \end{picture}
}
\end{center}
\caption{2PI Hartree vacuum diagram (left) and the one-loop self-energy diagram (right) contributing to the effective action $\Gamma_2$ and to the equations of motion, respectively.}
\label{fig:Hartree}
\end{figure}
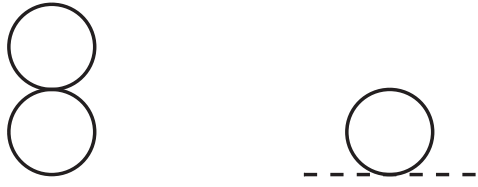

A way around this is to use a different "free" propagator to expand around (as in screened perturbation theory), or by carefully selecting a (infinite) sub-set of diagrams to re-sum, in order to dynamically generate a mass in a self-consistent way. One very popular such resummation is the Hartree approximation, which includes a single local diagram (see Fig.\ref{fig:Hartree}) in the propagator equation. The prescription is that the line in the diagram loop is itself the solution to the propagator equation, hence rendering the propagator self-consistent. The Hartree approximation is equivalent to resumming all Daisy and SuperDaisy diagrams, and thereby dynamically generating an effective mass. 

The Hartree approximation is the simplest case of a truncation of the 2PI-loop expansion for the effective action and all the 2PI $n$-point functions can be shown to be renormalizable \cite{Berges:2005hc}.  At the level of the action, it amounts to including the 2-loop "figure-8" vacuum diagram; at the level of the equation of motion, it amounts to including a local mass insertion proportional to the equal-time propagator in both mean field and propagator equations. Then both the mean field equation (as before) and the propagator equation need to be solved consistently. For a recent review of the 2PI technique, see \cite{Berges:2004yj} and references therein.

The standard expression for the 2PI effective action is
\ee{
\label{2PI_action}
\Gamma_{\rm 2PI}[\varphi,G,g^{\mu\nu}] = S_g[g^{\mu\nu}] + S_m[\varphi,g^{\mu\nu}] + \frac{i}{2}{\rm Tr}\ln G^{-1} + \frac{i}{2}{\rm Tr}\left[G_0^{-1} G\right]+\Gamma_2[\varphi,G,g^{\mu\nu}],}
where the free propagator is defined from
\ee{
iG_0^{-1}(x,y) = \frac{\delta S_m[\varphi,g^{\mu\nu}]}{\delta\varphi(x)\delta\varphi(y)}
= -\sqrt{-g}\left( -\square_y + m^2_0 + \xi_0 R + \frac{\lambda_0}{2}\varphi^2 \right)\delta(x-y).}
The form of $\Gamma_2[\varphi,G,g^{\mu\nu}]$ depends on the approximation used and our choice will be to use the first non-trivial approximation, two loops in the action, also known as the Hartree approximation. Hence we write
\ee{\Gamma_2[\varphi,G,g^{\mu\nu}]=- \frac{\lambda}{8}\int d^n x \sqrt{-g}\,G(x,x)^2.}
For our action defined via (\ref{eq:actm}) the 2PI action is
\ea{
\label{2PI_action2}
\Gamma_{\rm 2PI}[\varphi,G,g^{\mu\nu}] &= S_g[g^{\mu\nu}] -\f{1}{2}\int d ^nx\sqrt{-g}~ \bigg[\partial_\mu\varphi\partial^\mu\varphi+m^2_0\varphi^2 +\xi_0 R\varphi^2+2\f{\lambda_0}{4!}\varphi^4\bigg] + \frac{i}{2}{\rm Tr}\ln G^{-1}
 \nonumber \\ &- \frac{1}{2}\int d^nx\sqrt{-g}~\left(\nabla_{x,\mu}\nabla_y^\mu + m^2_1 + \xi_1 R + \frac{\lambda_1}{2}\varphi^2 \right) G(x,y)\bigg\vert_{x\rightarrow y}\nonumber\\
& - \frac{\lambda_2}{8}\int d^n x \sqrt{-g}\,G(x,x)^2\nonumber \\[2mm]
&\equiv S_g[g^{\mu\nu}]+\Gamma_{{\rm 2PI},m}[\varphi,G,g^{\mu\nu}],}
where following \cite{Arai:2012sh} we have explicitly written different bare couplings for each contribution in the 2PI action. In contrast to the 1PI case, we now also have an equation of motion for the propagator. All three equations can be derived via variations
\ee{
\frac{\delta \Gamma_{\rm 2PI}[\varphi,G,g^{\mu\nu}]}{\delta \varphi(x)} = 0,\qquad\frac{\delta \Gamma_{\rm 2PI}[\varphi,G,g^{\mu\nu}]}{\delta g^{\mu\nu}} = 0,\qquad
\frac{\delta \Gamma_{\rm 2PI}[\varphi,G,g^{\mu\nu}]}{\delta G(x,y)} = 0. 
\label{eq:vars2}}

\subsection{2PI Gap equation}
\label{sec:gap}

Next we will solve the propagator equation of motion. The equations of motion for the mean field and for the propagator from (\ref{eq:actm}) are 
\begin{align}
\label{eom_field_2PI}
&\left[-\square + m^2_0 + \xi_0 R
 + \frac{\lambda_0}{6}\varphi^2 + \frac{\lambda_1}{2}
G(x,x) \right] \varphi = 0,
\\
\label{eom_mode_2PI}
&\left[-\square_x + m^2_0 + \xi_0 R
 + \frac{\lambda_1}{2}\varphi^2 + \frac{\lambda_1}{2}
G(x,x) \right]G(x,y) =-i\f{\delta(x-y)}{\sqrt{-g}},
\end{align} 
where we used the fact that all the counter terms can be chosen to be equal, except for the $\delta\lambda_i$, for which we have \cite{Arai:2012sh}
\ee{
{\delta}\lambda_0 = 3{\delta} \lambda_1,}
with
\ee{
\lambda_0  =  \lambda+\delta\lambda_0 ,\qquad
\lambda_1 =  \lambda+\delta\lambda_1,}
such that all the counter terms have the property $c_i=c+\delta c_i$. The quantity of interest in this approximation is the self-consistent effective mass defined by equation (\ref{eom_mode_2PI}) as
\ee{
\label{eff_mass_2PI}
M^2_{\rm 2PI} \equiv m^2_0 + \xi_0 R
 + \frac{\lambda_1}{2}\varphi^2 + \frac{\lambda_1}{2}
G(x,x).}
If, as in section \ref{sec:1pisol}, we assume that $M_{\rm 2PI}$ is approximately constant it is easy to show that a mode satisfying (\ref{eq:fielddef}), (\ref{eq_com}) and (\ref{eq:sol}) satisfies equation (\ref{eom_mode_2PI}) if we simply replace $M$ with $M_{\rm 2PI}$. Deriving the 2PI counter terms in the Hartree approximation is a standard calculation which can be found in appendix \ref{sec:GapA}, see also \cite{Arai:2012sh}. By using the counter terms in (\ref{eq:count2pi}) and then setting $n=4$, from (\ref{eq:M2PI}) we can now straightforwardly derive the gap equation for the effective mass in (\ref{eff_mass_2PI}) with (\ref{eq:fullooppi})
\ee{M^2_{\rm 2PI}=\tilde{M}^2+\bigg(\frac{3\tilde{\lambda}}{16\pi ^2}\bigg)\frac{H^2}{M^2_{\rm 2PI}/H^2-3\epsilon+3 \epsilon^2+\delta_H\epsilon},\label{eq:gap}}
where we have defined $\tilde{M}^2=\tilde{m}^2+\tilde{\xi}R+\f{\tilde{\lambda}}{2}\varphi^2$ and the running constants 
\ee{\tilde{\lambda}=\frac{\lambda }{1-\frac{\lambda }{16 \pi ^2} \log\frac{H}{\mu'}},
\qquad\tilde{m}^2=\frac{m^2 }{1-\frac{\lambda }{16 \pi ^2} \log\frac{H}{\mu'}},
\qquad\tilde{\xi}=\f{1}{6}+\frac{\xi-\f{1}{6}}{1-\frac{\lambda }{16 \pi ^2} \log\frac{H}{\mu'}}.\label{eq:run}}
The solution of the gap equation (\ref{eq:gap}) is
\ee{\delta_{\rm 2PI} \equiv \f{M_{\rm 2 PI}^2}{H^2}=\frac{\tilde{\theta}}{2}+\sqrt{\frac{\tilde{\theta}^2}{4} + \frac{3\tilde{\lambda}}{16\pi ^2}} + 3\epsilon -3 \epsilon^2-\delta_H \epsilon\,,\label{eq:m2p1}}
with $\tilde{\theta}\equiv \tilde{M}^2/H^2 - 3\epsilon +3 \epsilon^2 +\delta_H \epsilon$.
Having solved for the effective mass, the renormalized field equation of motion (\ref{eom_field_2PI}) now reads
\ee{\ddot{\varphi}+3H\dot{\varphi}+ M_{\rm 2 PI}^2\varphi -\f{\lambda}{3}\varphi^3 = 0 \label{eq:feom2pi}}
We see that the infrared pole $1/\theta$ in the 1PI field equation (\ref{eq:eomfin1}), discussed in section \ref{sec:quant}, is lifted in the 2PI case due to self-consistent solution of the dynamical mass $M_{\rm 2 PI}$. The same observation was made for strict de Sitter space in refs.~\cite{Serreau:2011fu,Garbrecht:2011gu,Garbrecht:2013coa}, which agree with the result (\ref{eq:m2p1}) in the limit $\epsilon \to 0$.
In the 1PI limit with perturbative quantum corrections, given by (\ref{1PI_limit}), we get for the effective mass 
\ea{ M_{\rm 2 PI}^2\approx M^2+\f{\lambda H^2}{16\pi^2}\bigg\{(\delta+\epsilon-2)\log\left(\frac{H}{\mu'}\right)+\frac{3}{\theta}
\bigg\}\label{eq:app M}}
and therefore equation (\ref{eq:feom2pi}) reduces to the 1PI field equation (\ref{eq:eomfin1}), as expected.\footnote{The difference in scales ($\mu^{\prime}$ vs. $\mu^{\prime\prime}$) when comparing the above expression with (\ref{eq:finloop}) is due to $\mathcal{O}(n-4)$ term in the choice of $\delta\xi_0$ in (\ref{eq:count2pi}), which is irrelevant for 1PI according to the discussion in section \ref{sec:line}.}

\subsection{2PI Energy-Momentum tensor and Friedmann equations}
\label{2PI_energy}

By variation we can derive the energy-momentum tensor from (\ref{2PI_action})
\ea{
\label{EM-tensor_2PI}
T_{\mu\nu}^{\rm 2PI} &= -\frac{2}{\sqrt{-g}}\frac{\delta \Gamma_{{\rm 2PI},m}[\varphi,G,g^{\mu\nu}]}{\delta g^{\mu\nu}}\\
 &= \partial_\mu\varphi \partial_\nu\varphi -\frac{g_{\mu\nu}}{2}\Big(\partial_\rho\varphi \partial^\rho\varphi
  + m^2_0 \varphi^2 + 2\frac{\lambda_0}{4!}\varphi^4\Big) +\langle {\hat{T}^{Q}_{\mu\nu}}\rangle_*\nonumber\\&+g_{\mu\nu}\f{\lambda_2}{8}G^2(x,x)+\f{g_{\mu\nu}}{2}\xi_0 R G(x,x)+ \xi_0\Big( G_{\mu\nu} - \nabla_\mu\nabla_\nu + g_{\mu\nu}\square\Big)(\varphi^2+G(x,x)),\nonumber}
where $\langle\hat{T}^{Q}_{\mu\nu}\rangle_*$ denotes the one-loop energy-momentum tensor defined in (\ref{eq:energydensity}) 
with $M$ replaced by $M_{\rm 2PI}$ defined in (\ref{eq:m2p1}) and without the explicitly $\xi$ dependent piece. In order to find an explicit result for the energy-momentum, we can use (\ref{eff_mass_2PI}), (\ref{eq:energydensity}) and the counter terms from (\ref{eq:count2pi}) to get after some algebra
\ea{
T_{\mu\nu}^{\rm 2PI}&= -g_{\mu\nu}\bigg[\frac{1}{2} \partial_\rho\varphi\partial^\rho\varphi -\frac{\lambda}{12}   \varphi^4-\frac{M_{\rm 2PI}^2 \left( M_{\rm 2PI}^2-2 M^2\right)}{2 \lambda }\bigg]\nonumber \\&+\partial_\mu\varphi\partial_\nu\varphi +2\frac{\xi }{ \lambda }\bigg[ R_{\mu\nu}-\nabla_\mu\nabla_\nu+g_{\mu\nu}\Box\bigg]M_{\rm 2PI}^2\nonumber\\&-g_{\mu\nu}\frac{H^4}{32 \pi ^2}\left[\frac{6 \delta_{\rm 2PI}  }{ \delta_{\rm 2PI} -3 \epsilon +\delta_H\epsilon +3\epsilon^2}+ \left(\delta_{\rm 2PI} ^2-2 \delta_{\rm 2PI} -6\epsilon \right)\log\left(\frac{H}{\mu }\right)\right],\label{eq:T2PI}}
where we have neglected terms that are proportional to the gravitational counter terms in (\ref{eq:cters2}), and hence can be absorbed in them. Covariant conservation of (\ref{eq:T2PI}) is consistent with the field equation of motion (\ref{eq:feom2pi}) as is shown in appendix \ref{sec:appcov}. Taking the 1PI limit 
(\ref{eq:app M}) and expanding (\ref{eq:T2PI}) to 1-loop order we find agreement with the 1PI 1-loop results in section \ref{sec:1PI}. The surprising thing is that there is no need for any gravitational counter terms for removing the divergences, as (\ref{eq:count2pi}) are enough to render the energy-momentum tensor finite. We can simplify the above expression by using the gap equation (\ref{eq:gap}) and again ignoring terms that vanish after renormalization, which gives
\ee{T^{\rm 2PI}_{\mu\nu }=-\f{g_{\mu\nu}}{2} \partial_\rho\varphi\partial^\rho \varphi+\partial_\mu\varphi\partial_\nu\varphi+ \frac{2\xi }{\lambda}\Big [R_{\mu\nu}-(\nabla_\mu\nabla_\nu-g_{\mu\nu}\Box) \Big]M^2_{\rm 2PI}-g_{\mu\nu}W_{\rm 2PI}(\varphi,H,\epsilon)\label{eq:simp2PIem}}
where we have defined the potential
\ee{W_{\rm 2PI}(\varphi,H,\epsilon)\equiv-\frac{\lambda }{12}\varphi^4+\frac{M_{\rm 2PI}^4}{2 \tilde{\lambda}}+\bigg(\f{1}{\lambda}-\f{1}{\tilde{\lambda}}\bigg){ M_{\rm 2PI}^2H^2}+\frac{3\epsilon H^4}{\tilde{\lambda}}.\label{eq:simple2piT}}
We note that this potential differs from the ``true'' 2PI effective potential $V_{\rm 2PI}$, which gives the non-kinetic part of the field equation (\ref{eq:feom2pi}), defined as in (\ref{eq:feom2pi_2}).
By using (\ref{eq:simp2PIem}-\ref{eq:simple2piT}) we can again derive the Friedmann equations from (\ref{eq:nonrE})
\ea{3H^2&=\f{1}{M_{\rm pl}^2} \bigg[\f{1}{2}\dot{\varphi}^2+6\f{\xi}{\lambda}(H\partial_t-H^2)M_{\rm 2PI}^2+W_{\rm 2PI}(\varphi,H,\epsilon)\bigg]\label{eq:2piFr1}\\
a^2(-3H^2+2\epsilon H^2)&=\f{a^2}{M_{\rm pl}^2} \bigg[\f{1}{2}\dot{\varphi}^2+6\f{\xi}{\lambda}\bigg(-\f{1}{3}(2H\partial_t+\partial_t^2)+H^2\bigg)M_{\rm 2PI}^2-W_{\rm 2PI}(\varphi,H,\epsilon)\bigg].\label{eq:2piFr2}}

\subsection{2PI Renormalization conditions}
\label{sec:conditions2PI}
For 2PI we will use the same procedure as for 1PI in section \ref{sec:conditions} to impose the renormalization conditions and fix the finite parts of the counter terms.\footnote{We again split all constants to a physical part and a finite counter term: $c=c_{\textbf{ph}}+\tilde{\delta}c$.} We denote the renormalization point by (\ref{eq:incond}) and write the field equation (\ref{eq:feom2pi}) in terms of 2PI effective potential $V_{\rm 2PI}(\varphi,H,\epsilon)$ as
\ee{\ddot{\varphi}+3H\dot{\varphi}+\f{\partial V_{\rm 2PI}(\varphi,H,\epsilon)}{\partial \varphi}=0.
\label{eq:feom2pi_2}}
As before, our prescription is to renormalize $V_{\rm 2PI}(\varphi,H,\epsilon)$ to match the classical potential (\ref{class_potential}) at the renormalization point $\mu_0$: 
\ea{\f{\partial V_{\rm 2PI}(\varphi,H,\epsilon)}{\partial{\varphi}} \bigg\vert_{\mu_0}&=m^2_{\textbf{ph}} {\varphi_0 }+\frac{\lambda_{\textbf{ph}}  {\varphi_0 }^3}{6},&\f{\partial^2  V_{\rm 2PI}(\varphi,H,\epsilon)}{\partial\varphi^2}\bigg\vert_{\mu_0}&=m^2_{\textbf{ph}} +\frac{\lambda_{\textbf{ph}}  {\varphi_0 }^2}{2}, \nonumber \\\f{\partial^3V_{\rm 2PI}(\varphi,H,\epsilon)}{\partial\varphi^3}\bigg\vert_{\mu_0}&=\lambda_{\textbf{ph}}\varphi_0,&\f{\partial^4V_{\rm 2PI}(\varphi,H,\epsilon)}{\partial\varphi^4}\bigg\vert_{\mu_0}&=\lambda_{\textbf{ph}},\nonumber \\\f{\partial^4 V_{\rm 2PI}(\varphi,H,\epsilon)}{\partial H^2\partial \varphi^2}\bigg\vert_{\mu_0}&=0.\label{eq:finrenorm_2pi}}
Moreover, the same arguments regarding the renormalization of the cosmological constant and the Einstein tensor
give us the identical equations as in the 1PI case for the case of the minimal coupling $\xi_{\textbf{ph}}=0$:\footnote{For $\xi_{\textbf{ph}}=0$ the counter term $\delta\xi$ is a pure quantum contribution, whereby the correction to equation (\ref{eq:newfried_2pi}) of order $\delta\xi\,\epsilon\,\delta_{\rm 2PI}$ is negligible in our approximation.} 
\ea{3H_0^2& =\f{1}{M_{\rm pl}^2}\bigg(\f{1}{2}\dot{\varphi}_0^2+V_C(\varphi_0)\bigg)\label{eq:inE_2pi},\\
\epsilon H^2&=\f{\dot{\varphi}^2}{2 M_{\rm pl}^2}.\label{eq:newfried_2pi}}
In contrast to the 1PI case, however, the counter terms are now involved in both the tree-level \textit{and} loop contributions i.e the split $c=c_{\textbf{ph}}+\tilde{\delta}c$ must be performed for all constants in the potential $V_{\rm 2PI}(\varphi,H,\epsilon)$. This makes the analytical solution of the finite counter terms $\tilde{\delta}c$ complicated, but the numerical solution of (\ref{eq:finrenorm_2pi}) for the specific values of the renormalization point $\mu_0$ and the physical parameters $m_\mathbf{ph}$ and $\lambda_\mathbf{ph}$ is straightforward. 

\section{Quantum corrections to slow-roll parameters}
\label{sec:num}

In this section we will use the results of the previous sections to compute quantum corrections to slow-roll parameters $\epsilon$ and $\delta_H$, which contribute to the observables of the primordial power spectrum, in the minimally coupled case with $\xi_\mathbf{ph} = 0$, and estimate their size for both the massive and the massless $\varphi^4$ theory of inflation. In the end, we will also briefly comment on the role of quantum corrections in the curvaton scenario.  

Since we are not considering the full coupled dynamics of metric and matter field perturbations in this work, we do not attempt to derive the loop-corrected expression for the power spectrum of primordial curvature perturbations in the scalar-field inflation scenario (cf. \cite{Weinberg:2005vy,Weinberg:2006ac}). Instead, we use the classical result (cf. \cite{Liddle:2000cg,Weinberg:2008zzc}),
\ee{
P_{\cal R}(k) = \left(\frac{H}{\dot\varphi}\right)^2 \left(\frac{H}{2\pi}\right)^2\bigg|_{k=aH}\,,
\label{power_spec}
}
valid in the leading order of the slow-roll approximation. The expression (\ref{power_spec}) was also used in
 \cite{Bilandzic:2007nb}, where it was conjectured to be valid even in the presence of loop corrections, which, as far as we know, however has not been proved so far.

Assuming (\ref{power_spec}), the spectral index of nearly scale invariant perturbations is given by
\ee{
n_S(k) - 1 = \frac{\ln P_R(k)}{\ln(k)}\bigg|_{k=aH} = -4\epsilon_* - 2\delta_{H*}\,,
\label{spec_ind}
}
where we denote $\epsilon_* \equiv \epsilon|_{k=aH}$ and similarly for $\delta_H$. Likewise, for the gravitational wave spectrum we use the standard result:
\ee{
P_g(k) = \frac{8}{M_{\rm pl}^2}\left(\frac{H}{2\pi}\right)^2\bigg|_{k=aH}\,,
\label{tensor_spec}
}
such that the tensor-to-scalar ratio of the perturbations is given by
\ee{r \equiv \f{P_g}{P_{\cal R}} = \frac{8}{M_{\rm pl}^2}\left(\frac{\dot\varphi}{H}\right)^2\bigg|_{k=aH}
= 16\epsilon_* \,,
\label{tensor-scalar_ratio}}
where the last equality follows from (\ref{eq:newfried}).

To proceed, we derive formal expressions for $\epsilon$ and $\delta_H$ in terms of the effective
 potential $V$ and $H$. 
Within the slow-roll approximation, the second derivative $\ddot\varphi$ can be neglected in the
 field equation (\ref{scalar_eq_gen}) to give $\dot\varphi = -\partial_\varphi V / (3H)$, such
 that by (\ref{eq:newfried}) and (\ref{dot_eps}) we find
\ee{
\epsilon = \frac{\left(\partial_\varphi V\right)^2}{18 M_{\rm pl}^2 H^4},
\label{eps}
}
and 
\ee{
\delta_H 
 = \epsilon - \frac{\partial_\varphi^2 V}{3H^2}
 - \frac{\partial_\varphi V}{18 M_{\rm pl}^2 H^4}\Big[ H \partial_H \partial_\varphi V - 2(\epsilon + \delta_H)
\partial_\epsilon \partial_\varphi V\Big]\,.
\label{delH}
}   
In deriving (\ref{delH}) we also used the time-derivative of (\ref{eq:newfried}), again dropping
 the original $\ddot\varphi$ in accordance with the slow-roll approximation. Formally identical equations
 (\ref{eq:feom2pi_2},\ref{eq:newfried_2pi}) apply to the 2PI case as well, and hence
 the generic expressions (\ref{eps}-\ref{delH}) are valid for both 1PI and 2PI cases with $V$ and $M^2$ denoting
 the respective (1PI or 2PI) effective potential and dynamical mass. In general, however, the RHS's of 
(\ref{eps}-\ref{delH}) are fairly
 complicated functions of $\epsilon$ and $\delta_H$, and the exact analytical solution of these
 (algebraic) equations would be difficult. In what follows, we solve iteratively for the leading quantum
 corrections to $\epsilon$ and $\delta_H$ starting from their tree-level expressions. The error of this
 solution is of second order in the quantum corrections, parametrized by the dimensionless factors in
 Eqs.~(\ref{Q-correction1}-\ref{Q-correction11}) below.\footnote{More precisely, the error of an iterative
 solution $\epsilon_i$ to equation (\ref{eps}), written as $\epsilon = f(\epsilon)$, can be estimated by
 $|\epsilon - \epsilon_1| \leq |f^\prime(\epsilon_0)| |\epsilon_1 - \epsilon_0|$, assuming
 $|f^\prime(\epsilon_0)| \ll 1$, where $\epsilon_0$, $\epsilon_1$ and $\epsilon$ correspond to the zeroth
 iteration (classical solution), first iteration (classical + leading quantum correction) and the true
 solution, respectively. By direct computation we now find that the the size of the derivative 
 $f^\prime(\epsilon_0)$ is roughly controlled by the dimensionless factors in
 Eqs.~(\ref{Q-correction1}-\ref{Q-correction11}) and since the size of the leading correction
 $|\epsilon_1 - \epsilon_0|$ is controlled by the same factors, we conclude that the error
 $|\epsilon - \epsilon_1|$ is of second order in these factors. Essentially the same conclusion holds for
 the iterative solution of $\delta_H$.}

\subsection{Tree-level expressions}
\label{sec:tree}

At tree-level the scalar field effective potential is the classical one (\ref{class_potential})
 and the Friedman equation (\ref{eq:Fried_fin1}) reduces to
\ee{H_C^2 = \frac{T_{00}^C}{3M_{\rm pl}^2} \approx \frac{V_C}{3M_{\rm pl}^2}\,,
\label{eq:Fried_class1}
}
where we have neglected the kinetic term $\dot\phi^2/2$ in the energy density $T_{00}^C$ in accordance with the
 slow-roll approximation. By using (\ref{eq:Fried_class1}) and (\ref{class_potential}) we get for the slow-roll
 parameters (\ref{eps}-\ref{delH}) and $\delta = M^2/H^2$ at tree-level
\ea{
\epsilon_C &= \f{M_{\rm pl}^2}{2}\left( \f{\partial_\varphi V_C}{V_C}\right)^2
 = \left( \f{m_\mathbf{ph}^2 + \frac{\lambda_\mathbf{ph}}{6}\phi^2}{m_\mathbf{ph}^2
 + \frac{\lambda_\mathbf{ph}}{12}\phi^2}\right)^2 \f{2 M_{\rm pl}^2}{\phi^2}
\label{epsC}
\\
\delta_C &= 3 M_{\rm pl}^2 \f{\partial_\varphi^2 V_C}{V_C}
 = \left( \f{m_\mathbf{ph}^2 + \frac{\lambda_\mathbf{ph}}{2}\phi^2}{m_\mathbf{ph}^2
 + \frac{\lambda_\mathbf{ph}}{12}\phi^2}\right)\f{6 M_{\rm pl}^2}{\phi^2}
\label{delC}
}
and
\ee{
\delta_{HC} = \epsilon_C - \f{1}{3}\delta_C\,.
\label{delHC}
}
Similarly, the number of $e$-foldings from the end of inflation is given by 
\ee{
N_C = \int_t^{t_{\rm end}} H_C\,dt' = \frac{1}{8 M_{\rm pl}^2}\left(\varphi^2 -\varphi^2_{\rm end}
 + \frac{6m_\mathbf{ph}^2}{\lambda}\log\frac{m_\mathbf{ph}^2 + \f{\lambda_\mathbf{ph}}{6}\varphi^2}
{m_\mathbf{ph}^2 + \f{\lambda_\mathbf{ph}}{6}\varphi_{\rm end}^2}\right)\,.    
\label{NC}
}
where $\varphi_{\rm end}$ is obtained from the condition $\epsilon_{C,{\rm end}} = 1$.

\subsection{1PI quantum corrections}

By splitting the effective potential in (\ref{scalar_eq_gen}) into classical and quantum part:
 $V = V_C + V_Q$, with the similar split for the energy density given by
 (\ref{full_energy}-\ref{full_energyQ}), and using the tree-level expressions (\ref{class_potential}),
 (\ref{eq:Fried_class1}-\ref{delHC}) inside the quantum (loop) contributions, we get for the slow-roll
 parameters from (\ref{eps}-\ref{delH})
\ea{
\epsilon &= \epsilon_C + \epsilon_Q
\\
\delta_H &= \delta_{HC} + \delta_{HQ}
}
where the leading quantum corrections are given by
\ea{
\epsilon_Q =& \Bigg[\frac{2}{M_\mathbf{ph}^2\varphi - \frac{\lambda_\mathbf{ph}}{3}\varphi^3}
\left(\Delta M^2\varphi - \frac{\Delta\lambda}{3}\varphi^3 + \Delta\sigma + \f{1}{2}\Delta\eta\varphi^2
+ \frac{3\lambda_\mathbf{ph}\varphi H_C^2}{16\pi^2 \theta_C}\right) 
\nonumber\\
&-\f{2}{V_C}\left(\Delta\Lambda + \Delta V + \frac{3 H^4 \delta_\mathbf{ph} }{16 \pi ^2 \theta_C}\right)
\Bigg]\epsilon_C
\label{epsQ}
\\[4mm]
\delta_{HQ} =& -\frac{2}{M_\mathbf{ph}^2 - \frac{\lambda_\mathbf{ph}}{3}\varphi^2}\Bigg[\Delta\xi R_C
 + \frac{3\lambda_\mathbf{ph} H_C^2}{16\pi^2 \theta_C}\left(3 + \f{5\epsilon_C \delta_{HC} +
\delta_{HC}^2}{\theta_C}\right)\Bigg]\epsilon_C
\nonumber\\
&-\Bigg[\f{1}{M_\mathbf{ph}^2}\left(\Delta M^2 + \Delta\eta\varphi
 + \frac{3\lambda_\mathbf{ph} H_C^2}{16\pi^2 \theta_C}\left(1 - \frac{\lambda_\mathbf{ph}\varphi^2}{H_C^2 \theta_C}
\right)\right) 
\nonumber\\
&-\f{1}{V_C}\left(\Delta\Lambda + \Delta V + \frac{3 H^4 \delta_\mathbf{ph} }{16 \pi ^2 \theta_C}\right)
\Bigg]\f{\delta_C}{3}\,,
\label{delHQ}
}  
where $\theta$ is defined in (\ref{theta_def}), $\Delta V$ in (\ref{DelV}) and we denote
 $\Delta\Lambda \equiv - T_{00}^Q\big|_{\mu_0}$
 and $\Delta M^2 \equiv \Delta m^2 + \Delta\xi R + \f{\Delta\lambda}{2}\varphi^2$. The various $\Delta$'s
 arise from fixing the renormalization conditions (\ref{eq:finrenorm}) and (\ref{Lambda_finrenorm}),
 while the remaining terms include the ``direct'' loop contributions. The quantum corrections to the spectral index (\ref{spec_ind}) and the tensor-to-scalar ratio (\ref{tensor-scalar_ratio}) are then trivially given by
\ee{
\left(n_S - 1\right)_{Q*} = -4\epsilon_{Q*} - 2\delta_{HQ*}\,,\qquad\quad r_{Q*} = 16\epsilon_{Q*}\,.
\label{spec_r_Q}
}

Next, we evaluate the size of quantum corrections in two opposite limits with either the quadratic (mass) term
or the quartic term dominating the potential for the physically viable single-field inflaton parameters. 
 In the former case, we choose $m_\mathbf{ph}^2 \sim 10^{-11} M_{\rm pl}^2$ and $\lambda_\mathbf{ph} \sim 10^{-15}$
 such that $\lambda_\mathbf{ph}\varphi^2 \ll m_\mathbf{ph}^2$
 for the physically interesting scales
 $N \lesssim 100$. Furthermore, in this limit we find that
 $\delta_C - 3 \epsilon_C \approx 3\lambda_\mathbf{ph} M_{\rm pl}^2/ (2m_\mathbf{ph}^2) \ll 3\epsilon_C^2
 \approx 3/(2N_C+1)^2$, such that the IR enhancement factor in the loop contributions gives
\ee{
\frac{1}{\theta_C} \approx \frac{1}{3\epsilon_C^2} \approx \frac{(2N_C + 1)^2}{3}\,.
}    
Therefore, since $H_C^2/M_\mathbf{ph}^2 \approx (2N_C + 1)/3$ and $H_C^4 \delta_\mathbf{ph}/V_C \approx m_\mathbf{ph}^2/(3M_{\rm pl}^2)$, we find for the size of $V_Q$- and $T_{00}^Q$-induced quantum corrections in (\ref{epsQ}-\ref{delHQ}), respectively
\ea{
\frac{3\lambda_\mathbf{ph}}{16\pi^2 \theta_C}\cdot\frac{H_C^2}{M_\mathbf{ph}^2} &\approx 
\frac{\lambda_\mathbf{ph}}{16\pi^2} \frac{(2N_C + 1)^3}{3},
\label{Q-correction1}
\\
\frac{3}{16\pi^2 \theta_C}\cdot\frac{H_C^4 \delta_\mathbf{ph}}{V_C} &\approx 
\frac{1}{16\pi^2} \f{m_\mathbf{ph}^2}{M_{\rm pl}^2}\f{(2N_C + 1)^2}{3}.
\label{Q-correction11}
}
We see that the coupling constant is enhanced by a large factor $(2N_C+1)^3/3$ due to the IR effects.
 However, for tiny coupling $\lambda_\mathbf{ph} \sim 10^{-15}$ the size of the correction (\ref{Q-correction1})
 is totally negligible for the physically interesting scales $N \lesssim 100$. For the EM-tensor-induced
 quantum corrections (\ref{Q-correction11}) the coupling constant $\lambda_\mathbf{ph}$ is replaced by $m_\mathbf{ph}^2/M_{\rm pl}^2 \sim 10^{-11}$ while the enhancement factor is reduced by $2N_C +1$ in comparison with (\ref{Q-correction1}), resulting in slightly larger but negligible corrections. 

In the latter case, we choose $m_\mathbf{ph} = 0$ and $\lambda_\mathbf{ph} \sim 10^{-12}$ to find for 
the IR enhancement factor
\ee{
\frac{1}{\theta_C} \approx \frac{2}{3\epsilon_C} \approx \frac{2(N_C + 1)}{3}\,,
}    
while $H_C^2/M_\mathbf{ph}^2 \approx 2(N_C + 1)/9$. Hence we get for the size of the $V_Q$-induced quantum corrections
\ee{
\frac{3\lambda_\mathbf{ph}}{16\pi^2 \theta_C}\cdot\frac{H_C^2}{M_\mathbf{ph}^2} \approx 
\frac{\lambda_\mathbf{ph}}{16\pi^2} \frac{4(N_C + 1)^2}{9}\,,
\label{Q-correction2}
}
with a similar expression for the $T_{00}^Q$-induced corrections, ie. the coupling constant is now
 enhanced by $\sim (2N_C)^2/9$, one factor of $N_C$ less than in the previous case.
Also in this case the quantum corrections are negligible due to the smallness of 
$\lambda_\mathbf{ph}$. A similar conclusion was obtained in \cite{Bilandzic:2007nb}, where it was found,
 however, that for the minimally coupled case $\xi_\mathbf{ph} = 0$ the quantum corrections would be enhanced
 by just one power of $N$. We have not been able to track down the origin of this discrepancy explicitly. 

Finally, in both of these cases we find that $\theta_C \gg \sqrt{\lambda_\mathbf{ph}}/(4\pi)$ and therefore
 the non-perturbative SuperDaisy resummation performed in 2PI Hartree approximation should not give relevant
 corrections in comparison to the 1PI results presented here. 

\subsection{Curvaton}
In the curvaton scenario \cite{Linde:1996gt,Enqvist:2001zp,Lyth:2001nq,Moroi:2001ct} the primordial density perturbations are generated by an auxiliary field, the curvaton, of which the energy density is subdominant during inflation, but which
acquires quantum perturbations during inflation. After inflation, these
give rise to an almost scale-invariant spectrum of curvature perturbations with the spectral index
 \cite{Lyth:2001nq}
\ee{n_S-1=-2\epsilon+\f{2}{3}\delta,\label{eq:curv_spectrum}}
where the slow-roll parameters are defined as in (\ref{eq:defeps}) and (\ref{eq:defdel}). In a self-interacting curvaton scenario \cite{Enqvist:2010dt}, the quantum corrected 1PI equation of motion for the curvaton mean field is of the form (\ref{scalar_eq_fin}), where we now find for the IR denominator (\ref{theta_def}) appearing in the loop correction terms
\ee{
\theta 
\approx \f{2}{3}(n_S-1) \approx -0.027,
\label{eq:curv_spectrum2}
}
for the physical value \cite{Ade:2013uln} $n_S \approx 0.96$. However, a negative value for $\theta$ is incompatible with the evaluation of the loop integral in $G(x,x)$ as it would result in an IR-divergence, ie. the results (\ref{eq:full_loop}-\ref{eq:energydensity}) are valid {\em only} if $\theta > 0$. For this reason it seems doubtful if the 1PI approximation can be used to study the quantum corrections in this model,
although it is possible that the classical relation (\ref{eq:curv_spectrum}) may be altered significantly by the 1PI corrections to accommodate this discrepancy.

Perhaps more likely, since 
classically the curvaton field appears to be very light, $\delta < 3\epsilon$, a self-consistent resummation of the IR effects may give rise to non-perturbatively large quantum corrections to the dynamical mass, which could be studied in the 2PI Hatree approximation using the formalism of section \ref{sec:2PI1}. Nevertheless, it seems that the quantum corrections may have a significant effect in the curvaton dynamics and we feel that this calls for further investigations. 
  
\section{Conclusions}
\label{sec:conc}

In this work we have considered the dynamics of a massive $\varphi^4$ scalar field theory in slow-roll quasi-de Sitter Universe. We have computed one-loop quantum corrections to the field and Friedmann equations of motion to the leading order in the slow-roll expansion, both in the 1PI and the resummed 2PI expansion. We have renormalized the equations of motion with the slow-rolling vacuum state, ie. expanding around de Sitter rather than Minkowski vacuum. Using these results, we have computed leading quantum corrections to slow-roll parameters and estimated their effect on the primordial power spectrum.

Like in the de Sitter case, we have found that for a light field with $M^2 \ll H^2$ the leading quantum correction gets enhanced by the IR part of the loop integral, resulting in an effective enhancement factor
\ee{\frac{1}{\theta} \equiv \frac{1}{M^2/H^2-3\epsilon + 3 \epsilon^2+\epsilon\delta_H},
}       
where $\epsilon$ and $\delta_H$ are the first two slow-roll parameters. Due to the minus sign in front of $3\epsilon$ this factor is typically larger than the one in the strict de Sitter case with $\epsilon = 0$. If the IR enhancement is large enough that $\theta \lesssim \sqrt{\lambda}/(4\pi)$ the perturbative 1PI results cannot be trusted and a self-consistent resummation of the IR effects contributing to the dynamical mass of the field is required. At one-loop level this corresponds to SuperDaisy resummation of the propagator and it is accounted for in 2PI Hartree approximation. The resulting self-mass is given by (\ref{eq:m2p1}), which generalizes de Sitter results in \cite{Serreau:2011fu} (see also \cite{Gautier:2013aoa,Garbrecht:2011gu,Garbrecht:2013coa}) to quasi-de Sitter case and the 1PI results in \cite{Boyanovsky:2005sh} to leading order in 2PI.    

Moreover, we found that in the Friedmann equations at the leading order the same quantum correction enters in the energy density and the pressure, while the kinetic and gradient terms for the scalar field remain uncorrected. Therefore, in the minimally coupled case $\xi = 0$ we recovered the classical relation (\ref{eq:newfried}) between the time-derivatives of the mean field and the Hubble rate. For comparison, in \cite{Markkanen:2012rh} it was found that expanding up to four gradients around Minkowski space, the corrections to the Friedmann equations involve kinetic terms, and the new contributions in energy density and pressure are no longer the same. Also in the present case, however, the quantum correction to the potential force in the field equation does not follow from simple variation from the potential-like quantum contribution in the Friedmann equation, which prevents from using a full analogy with the classical slow-roll formalism.

Due to the smallness of the coupling $\lambda \lesssim 10^{-12}$ for the massive $\varphi^4$ theory of inflation, we found that the quantum correction to the slow-roll parameters and thereby to the primordial power spectrum are negligible despite the large IR enhancement by a factor proportional to up to third power in the number of $e$-foldings. However, as stated above, we have not considered the effect of quantized metric perturbations on the dynamics of the matter fields, which may have a significant effect on the loop corrections in the inflaton scenario. On the other hand, for a curvaton scenario we found tension in using the classical expression for the power spectrum together with 1PI loop corrections, which may indicate that the quantum corrections could be non-perturbatively large in this case, thus requiring resummation. It would therefore be interesting to study this scenario more carefully in 2PI Hartee approximation using the methods and results presented in this work.       
\section*{Acknowledgments}
We would like to thank Kari Rummukainen, Mark Hindmarsh and Sami Nurmi for useful and illuminating discussions. TM is funded by the doctoral programme in PArticle- and NUclear physics (PANU) of the University of Helsinki, and acknowledges support from The Academy of Finland through project number 1134018. MH and AT acknowledge support from the Villum Foundation.

\appendix
\section{Geometric tensors in $n$ dimensional FRW spaces}
\label{sec:appA}
In this section we present the expressions for the geometric tensors in a FRW space-time to first order in the slow-roll expansion. For the full expressions see \cite{Markkanen:2013nwa}.

Standard variational calculus gives the following geometric tensors
\begin{align}
G_{\mu\nu}\equiv\f{1}{\sqrt{-g}}\f{\delta}{\delta g^{\mu\nu}}\int d^nx\sqrt{-g}~R=-\f{1}{2} Rg_{\mu\nu}+R_{\mu\nu},
\end{align}
\begin{align}
\f{1}{\sqrt{-g}}\f{\delta}{\delta g^{\mu\nu}}\int d^nx\sqrt{-g}~Rf(x)=\big[-\f{1}{2} Rg_{\mu\nu}+R_{\mu\nu}-\nabla_\mu\nabla_\nu+g_{\mu\nu}\Box\big]f(x),
\end{align}
\begin{align}
~^{(1)}H_{\mu\nu}\equiv\f{1}{\sqrt{-g}}\f{\delta}{\delta g^{\mu\nu}}\int d^nx\sqrt{-g}~R^2=-\f{1}{2} R^2g_{\mu\nu}+2R_{\mu\nu}R-2\nabla_\mu\nabla_\nu R+2g_{\mu\nu}\Box R,
\end{align}
\begin{align}
~^{(2)}H_{\mu\nu}&\equiv\f{1}{\sqrt{-g}}\f{\delta}{\delta g^{\mu\nu}}\int d^nx\sqrt{-g}~R^{\mu\nu}R_{\mu\nu}\nonumber \\&=-\f{1}{2}R_{\alpha\beta}R^{\alpha\beta}g_{\mu\nu}+ 2R_{\rho\nu\gamma\mu}R^{\rho\gamma}-\nabla_\nu\nabla_\mu R+\f{1}{2}\Box R g_{\mu\nu}+\Box R_{\mu\nu},
\end{align}
and
\begin{align}
H_{\mu\nu}&\equiv\f{1}{\sqrt{-g}}\f{\delta}{\delta g^{\mu\nu}}\int d^nx\sqrt{-g}~R^{\mu\nu\sigma\delta}R_{\mu\nu\sigma\delta}\nonumber \\&=
-\f{g_{\mu\nu}}{2}R^{\alpha\sigma\gamma\delta}R_{\alpha\sigma\gamma\delta} +2{R_\mu}^{\rho\alpha\sigma}R_{\nu\rho\alpha\sigma}+4R_{\sigma\mu\gamma\nu}R^{\gamma\sigma}- 4R_{\mu\gamma}{R^\gamma}_{\nu}+4\Box R_{\mu\nu}-2\nabla_\mu\nabla_\nu R.
\end{align}
The second order tensors in a FRW universe are
\ea{(-\nabla_0\nabla_0+g_{00}\Box)f(t)&=(n-1)H\dot{f}(t), \\
(-\nabla_i\nabla_i+g_{ii}\Box)f(t)&=a^2\bigg[(2-n)H\dot{f}(t)-\ddot{f}(t)\bigg],\\
R&=(-1+n) (n-2 \epsilon ) H^2\label{eq:Rt},\\
G_{00}&=\f{(n-1)(n-2)}{2}H^2,\\
G_{ii}&=-\frac{1}{2}(-2+n)(-1+n-2 \epsilon ) a^2 H^2.}
Similarly, the fourth order tensors are given by
\ea{ ~^{(1)}H_{00}&=\frac{1}{2} (-1+n)^2 \left(n^2+8 \epsilon -4 n (1+2 \epsilon )\right) H^4,\\
 ~^{(1)}H_{ii}&=-\frac{1}{2} (-1+n) \left(n^3-8 \epsilon +4 n (1+8 \epsilon )-n^2 (5+12 \epsilon )\right) a^2 H^4,\\ 
 ~^{(2)}H_{00}&=-\frac{1}{2} (-1+n)^2 (4+n (-1+2 \epsilon )) H^4,\\ 
 ~^{(2)}H_{ii}&=\frac{1}{2} (-1+n) \left(n^2 (-1+2 \epsilon )+n (5+2 \epsilon )-4 (1+4 \epsilon )\right) a^2 H^4,\\ 
 H_{00}&=(-1+n) (-4+n+4 \epsilon -4 n \epsilon ) H^4,\\ 
 H_{ii}&=\left(-4+5 n-n^2+4 (-3+(-1+n) n) \epsilon \right) a^2 H^4.}

\section{The approximation for the loop integrals}
\label{sec:quant}

The objective of this work is to acquire expressions for the loop corrections that in addition to the ultraviolet contributions would contain the leading infrared terms. Indeed, for quite some time it has been known how to derive the ultraviolet, or in other words local, terms (cf. \cite{Markkanen:2012rh} and references therein), but there have been few attempts to incorporate also the infrared contributions for the case of a non-static space-time. 

First, we notice that our approximation for the mode functions, (\ref{eq:sol}), inherently neglects the time-derivative of the index $\nu$ of the Hankel functions. This is justified because $\dot\nu \sim {\cal O}(\epsilon^2, \epsilon\delta_H, \epsilon\delta )$ is of second order in the slow-roll parameters. For this reason, however, we have to be careful that the corresponding contributions (proportional to $\dot\nu$) are neglected when computing and manipulating the expressions for the loop corrections using the solution (\ref{eq:sol}).  

As we will show explicitly in section \ref{sec:loopEM}, the infrared integrals evaluate to terms proportional to 
\ee{\f{1}{3-2\nu}\approx\frac{3}{2 \big(\delta-3 \epsilon +3 \epsilon ^2+\delta_H\epsilon\big)}
+\cdots.\label{eq:nuapp}}
which we have expanded to leading order assuming $\epsilon \sim \delta \sim \delta_H$ with $\nu$ defined in (\ref{eq:def}). We have kept the second order terms in denominator as it may turn out that $\delta - 3\epsilon \lesssim 3\epsilon^2$, see section 4 below for an example.
  In our calculation we also encounter derivatives
\ee{\partial_t\f{1}{3-2\nu},\qquad(\partial_t)^2\f{1}{3-2\nu}\label{delte2},}
which are proportional to $\dot\nu$. Therefore, assuming the time-derivatives of $\epsilon$, $\delta$ and $\delta_H$ are sufficiently small, we indeed find that these contributions are subleading and can be neglected, consistent with the above prescription. For instance, using (\ref{dot_eps}) we find that the part proportional to $\dot\epsilon$ of the time-derivative of (\ref{eq:nuapp}) is suppressed by an additional power of $\epsilon$ (or $\delta_H$) when compared with (\ref{eq:nuapp}). For consistency, we also neglect the other IR and IM contributions of the same order or higher as (\ref{delte2}), eg. the terms proportional to (\ref{eq:nuapp}) multiplied by additional powers of $\epsilon$, $\delta$ or $\delta_H$.

On the other hand, for the UV contributions we also include those linear order (and even higher for some contributions) terms in $\epsilon$, $\delta$ and $\delta_H$ that are relevant for the renormalization. We notice, however, that in the second order the inherently neglected time-derivative $\dot\nu$ would become comparative and therefore cause imprecision. In section \ref{sec:line} we show that for 1PI the covariant linear terms in $\epsilon$ and $\delta$ can be discarded in the calculation by suitable redefinitions in renormalization. To summarize:
\begin{itemize}
\item We take the leading order loop contribution, which in our approximation arises from the infrared (IR) part of the loop integral and is proportional to the factor (\ref{eq:nuapp}).
\item
In addition, we include those linear order (or even higher) in $\epsilon$, $\delta$ and $\delta_H$ ultraviolet (UV) contributions that are relevant for the renormalization, ie. the UV divergences and the finite terms within the same structures.
\end{itemize}

\section{Calculation of the 1-loop variance and Energy-Momentum tensor}
\label{sec:loopEM}

In this section we will presents the details of the derivation of the results (\ref{eq:full_loop}) and (\ref{eq:energydensity}). 
Using the definitions (\ref{eq:fielddef}) and (\ref{eq_com}) the variance is simply
\ee{\langle \hat{\phi}^2\rangle=\int d^{n-1}\vert\mathbf{k}\vert\,\vert u_{\mathbf{k}}\vert^2.}
By changing the integration variable from $\vert\mathbf{k}\vert$ to 
\ee{x=\f{\vert\mathbf{k}\vert}{aH(1-\epsilon)},}
and using the solution (\ref{eq:sol}) for the mode functions we can write
\ea{\langle \hat{\phi}^2\rangle&=\f{\mu^{4-n}\sqrt{\pi}}{4\Gamma[\f{n-1}{2}]}\bigg(\f{(1-\epsilon)H}{2\sqrt{\pi}}\bigg)^{n-2}\int_0^\infty dx~x^{n-2}
\vert H_\nu^{(1)}(x)\vert^2
,}
where we have have 
introduced an arbitrary scale $ \mu^{4-n}$ to maintain the proper mass dimension of the variance. Following the steps of \cite{Serreau:2011fu}, we introduce dimensionless cut-off parameters $\kappa_{\rm IR}$ and $\kappa_{\rm UV}$ with the properties
\ee{\kappa_{\rm IR}\ll1\ll\kappa_{\rm UV},}
and split the integral into three regions as 
\ee{\int_0^\infty dx=\int_0^{\kappa_{\rm IR}}dx+\int_{\kappa_{\rm IR}}^{\kappa_{\rm UV}}dx+\int_{\kappa_{\rm UV}}^\infty dx,} which we call the infrared (IR), the intermediate (IM) and the ultraviolet (UV) regions, respectively. We will be using dimensional regularisation to regulate the UV divergencies, and since the divergences enter only in the UV region, we can set $n=4$ for the infrared and intermediate regions. The IR integral can be calculated by using the asymptotic expansions of the Hankel function
\ee{H_{\nu}^{(1)}(x)=-i\bigg(\f{2}{x}\bigg)^\nu\f{\Gamma[\nu]}{\pi}+\mathcal{O}(x^\nu)
.}
The integral can now straightforwardly be performed, giving
\ee{\langle \hat{\phi}^2\rangle_{\rm IR} \approx \frac{H^2}{4 \pi ^2 }\bigg(\f{1}{3-2 \nu }+\log\left(\kappa _{\text{IR}}\right)\bigg)
\label{eq:loopIR},}
where we have neglected the linear terms in $\epsilon$ and $\delta$ according to the discussion in section \ref{sec:quant}, as well as terms of order $\kappa _{\text{IR}}^2$. 
The IM integral can be calculated to $\mathcal{O}(1)$ accuracy by taking the limit $\epsilon,\delta \to 0$ with $\nu \to 3/2$ for the Hankel function and using the exact expression
\ee{H_{3/2}^{(1)}(x)=-\frac{e^{i x} \sqrt{\frac{2}{\pi }} (i+x)}{x^{3/2}},
}
to get
\ee{\langle \hat{\phi}^2\rangle_{\rm IM}=\frac{H^2 }{4 \pi ^2}\left[\log\left(\f{\kappa _{\text{UV}}}{\kappa _{\text{IR}}}\right)+\frac{\kappa _{\text{UV}}^2}{2}\right] + \mathcal{O}(\epsilon,\delta,\kappa _{\text{IR}}^2).
\label{eq:loopIM}}
In the UV region we will take advantage of the large-$x$ expansions of the Hankel function:
\ea{H^{(1)}_\nu(x)&=-\frac{e^{i \left(x-\frac{\pi  \nu }{2}\right)}}{\sqrt{\pi  x}}\left[{(1 - i) }+\frac{(1 +i) \left( \nu ^2-1/4\right)}{2 x}-\frac{(1 - i)\left(9-40 \nu ^2+16 \nu ^4\right)}{128x^2}+\mathcal{O}(x^{-7/2})\right]
.} 
Because of the UV divergencies we cannot set $n=4$ for this contribution. To evaluate the UV integral we
 will make use of the identity 
\ee{\int_{\kappa_{\rm UV}}^\infty dx~x^{n+\alpha-1}=-\f{({\kappa_{\rm UV}})^{n+\alpha}} {n+\alpha}.
\label{dim_int}}
The integral in (\ref{dim_int}) is convergent only for $Re[n + \alpha] < 0$, but the identity (\ref{dim_int}) can be defined as an analytic continuation for arbitrary complex values of $n+\alpha$. This is equivalent to assuming that the integral $\int_{0}^\infty dx\,x^{n+\alpha-1}$ would be vanishing for all values of $n+\alpha$, in accordance with the standard prescription of dimensional regularization.
The UV integral then gives
\ea{\langle \hat{\phi}^2\rangle_{\rm UV} \approx \frac{H^2}{8\pi ^2}\Bigg\{&(2-\delta  - \epsilon )\left[\frac{1}{4-n}-\log\left(\frac{H}{\mu }\right)\right]+\frac{1}{2} -\gamma_e+ \log(\pi)
-2 \log\left(\kappa _{\text{UV}}\right)-\kappa _{\text{UV}}^2 \Bigg\},
\label{eq:loopUV}}
where we have neglected the terms of second order and higher in $\epsilon$, $\delta$ and $\delta_H$, as well as terms of order $\kappa_{\text{UV}}^{-2}$. We have also neglected some other terms of order $\epsilon$, see below after (\ref{eq:fullooppi}).
Combining (\ref{eq:loopIR}), (\ref{eq:loopIM}) and (\ref{eq:loopUV}), we get for the entire loop contribution
\ee{\langle \hat{\phi}^2\rangle\approx\frac{H^2}{8\pi ^2}\left\{(-\delta  - \epsilon +2)\left[\frac{1}{4-n}-\log\left(\frac{H}{\mu' }\right)\right]+\frac{3}{ \delta -3\epsilon +\delta_H\epsilon+3\epsilon^2}
\right\},
\label{eq:fullooppi}}
where we have expanded the IR denominator using (\ref{eq:nuapp}) and the scale $\mu'$ is defined as
\ee{\mu'=\mu\text{ exp}\bigg[\f{1}{4}\Big(1 - 2  \gamma_e  + 2 \log(\pi)\Big)\bigg].\label{eq:mu}}
For the 1PI case $\mu'$ can be replaced by $\mu$ by using the arguments of section \ref{sec:line} regarding the terms of linear order in $\delta$ and $\epsilon$. Furthermore, in (\ref{eq:fullooppi}) and before in (\ref{eq:loopUV}) we have neglected UV terms of order $\epsilon$ and $\delta$, which do not combine with the divergence $1/(4-n)$ under the common factor $2-\delta-\epsilon$ and therefore are not directly related to renormalization, according to the discussion in section \ref{sec:quant}.

In the same way we can derive the result for the quantum energy-momentum tensor in (\ref{eq:quantumEM}). Since we already have the result for the variance in (\ref{eq:fullooppi}), we only need the contributions for the (explicitly) $\xi$-independent terms in the energy density
\ea{\langle\hat{T}^Q_{00}\rangle&=\f{1}{2}\bigg[\f{\partial}{\partial x_\rho}\f{\partial}{\partial y^\rho} +M^2 \bigg]G(x,y)\big\vert_{x=y}+\f{\partial}{\partial x^0}\f{\partial}{\partial y^0}G(x,y)\big\vert_{x=y} \nonumber \\&+\xi\big[R_{00}-\nabla_0\nabla_0-\Box\big]G(x,x) \nonumber \\&=\f{1}{2}\int d^{n-1} {k}\bigg[\vert\dot{u}_\mathbf{k}\vert^2+\bigg(\f{\mathbf{k}^2}{a^2}+M^2\bigg)\vert{u}_\mathbf{k}\vert^2+2\xi\big(R_{00}+ (n-1)H\partial_0\big)\vert{u}_\mathbf{k}\vert^2\bigg],}
to get the IR
\ee{~_{\xi=0}\langle\hat{T}^Q_{00}\rangle_{\rm IR}\approx\frac{\delta  H^4}{8 \pi ^2 (3-2 {\nu})},
\label{eq:Tir}}
the IM
\ee{~_{\xi=0}\langle\hat{T}^Q_{00}\rangle_{\rm IM}\approx\frac{H^4 \kappa _{\text{UV}}^2 \left(1+\kappa _{\text{UV}}^2\right)}{16 \pi ^2}
\label{eq:Tim},}
and the UV contributions
\ea{~_{\xi=0}\langle\hat{T}^Q_{00}\rangle_{\rm UV}&\approx\frac{H^4}{32 \pi ^2 }\left\{\left(-\delta ^2-4\delta\epsilon+2 \delta +6 \epsilon \right)\left[\frac{1}{4-n }-\log\left(\frac{H}{\mu }\right)\right]-3 -2 \kappa _{\text{UV}}^2-2 \kappa _{\text{UV}}^4\right\}.
\label{eq:Tuv}}
Combining the results in (\ref{eq:fullooppi}) and (\ref{eq:Tir} - \ref{eq:Tuv}) gives\ea{\langle\hat{T}^Q_{00} \rangle&=\frac{H^4}{32 \pi ^2 }\bigg\{6\frac{ \delta -6 \xi}{\delta -3 \epsilon+\delta_H\epsilon+3\epsilon^2 }-\Big(\delta ^2-2 \delta  (6 \xi +1)+24 \xi +\delta  (4-12 \xi ) \epsilon \nonumber \\&-6 (1-2 \xi ) \epsilon \Big)\left[\frac{1}{4-n}-\log\left(\frac{H}{\mu }\right)\right]\bigg\},\label{eq:energydensity2}}
up to higher order terms defined in the same way as for the variance (\ref{eq:fullooppi}), see the discussion in section \ref{sec:quant}. We have also discarded the contributions proportional to $H^4$, which can be removed by renormalization of $~^{(3)}H_{00}$ as defined in (\ref{eq:3H00}). Similarly, for the pressure we have 
\ee{~_{\xi=0} \langle\hat{T}^Q_{ii}\rangle=\f{1}{2}\int d^{n-1} {k}\bigg[\vert\dot{u}_\mathbf{k}\vert^2-\bigg(\f{3-n}{1-n}\f{\mathbf{k}^2}{a^2}+M^2\bigg)\vert{u}_\mathbf{k}\vert^2\bigg],}
to get
\ea{\langle\hat{T}^Q_{ii}\rangle/ a^2=-\langle\hat{T}^Q_{00}\rangle
\label{eq:pdens}.}
We have also checked that the same approximate expressions (\ref{eq:fullooppi}) and (\ref{eq:energydensity2}) are obtained by computing the loop integral in $n$-dimensional coordinate space, which is straightforward since in our approximation $\epsilon$, $\delta$ and $\delta_H$ are effectively constants in the loop integration and therefore $G(x,x)$ can be expressed in terms of the hypergeometric function ${}_2 F_1(a,b,c;z)$.

\subsection{1PI counter terms and finite results}
\label{Sec:C-terms}
Using the expressions (\ref{eq:fullooppi}) and (\ref{eq:energydensity2}) for $\langle \hat{\phi}^2\rangle$ and $\langle\hat{T}^Q_{00}\rangle$ it is a simple exercise of linear algebra to renormalize the 1PI equations of motion (\ref{eq:eom1}) and (\ref{eq:nonrE}). The divergent counter terms are given by
\ea{&{\delta} \xi =\frac{\lambda  (6 \xi -1) }{96 \pi ^2 (4-n)},&{\delta} \lambda &=\frac{3 \lambda ^2}{16\pi ^2 (4-n) }, &{\delta}m^2&=\frac{m^2 \lambda }{16 \pi ^2(4-n) },\nonumber \\ &{\delta}\Lambda=-\frac{m^4}{32\pi ^2 (4-n) },&{\delta}\alpha&=\frac{m^2 (1-6 \xi )}{96 \pi ^2 (4-n)},&{\delta}\beta&=\frac{-228 \xi ^2+20 \xi +3}{1152 \pi ^2 (4-n)} ,\nonumber \\&{\delta}\epsilon_1=-\frac{-48 \xi ^2+2 \xi +1}{96 \pi ^2 (4-n)}, &{\delta}\epsilon_2&=0,}
whereby the divergence in the variance (\ref{eq:fullooppi}) is cancelled and we are left with the finite
 result 
\ee{\langle \hat{\phi}^2\rangle_{\rm fin}=\frac{H^2}{8\pi ^2}\left\{(\delta  + \epsilon -2)\log\left(\frac{H}{\mu'' }\right)+\frac{3}{ \delta -3\epsilon+\delta_H\epsilon+3\epsilon^2 }\right\},\label{eq:finloop}}
where an additional finite correction resulting from the dimensional $4-n$ expansion of the Ricci scalar $R$ with the counter-term $\delta\xi$ is absorbed in the new scale $\mu''=e^{7/12}\mu'$. 
Furthermore, the results for the finite energy density and pressure are simply the expressions in (\ref{eq:energydensity2}) and (\ref{eq:pdens}) with the poles $1/(4-n)$ discarded, given by (\ref{eq:finEM}).

The $\Delta$-terms in the field equation (\ref{scalar_eq_fin}) induced by the renormalization conditions (\ref{eq:finrenorm}) fixing the finite parts of the counter terms are given by
\ea{\Delta \sigma&=\frac{\lambda_{\textbf{ph}} H_0^2  \varphi_0}{2 \pi ^2}\bigg[\frac{81 \epsilon_0^4}{\theta_0^4}-\frac{162 \epsilon_0^3}{\theta_0^4}+\frac{81 \epsilon_0^2}{\theta_0^4}-\frac{45 \epsilon_0^2}{\theta_0^3}+\frac{45 \epsilon_0}{\theta_0^3}+\frac{6}{\theta_0^2}+\frac{3 m^4_{\textbf{ph}}}{\theta_0^4 H_0^4} \nonumber \\&+\frac{81 H_0^2 \epsilon_0^6}{\theta_0^4 m^2_{\textbf{ph}}}-\frac{243 H_0^2 \epsilon_0^5}{\theta_0^4 m^2_{\textbf{ph}}}+\frac{243 H_0^2 \epsilon_0^4}{\theta_0^4 m^2_{\textbf{ph}}}-\frac{81 H_0^2 \epsilon_0^3}{\theta_0^4 m^2_{\textbf{ph}}}+\frac{27 m^2_{\textbf{ph}} \epsilon_0^2}{\theta_0^4 H_0^2}-\frac{27 m^2_{\textbf{ph}} \epsilon_0}{\theta_0^4 H_0^2}\nonumber \\&-\frac{135 H_0^2 \epsilon_0^4}{2 \theta_0^3 m^2_{\textbf{ph}}}+\frac{135 H_0^2 \epsilon_0^3}{\theta_0^3 m^2_{\textbf{ph}}}-\frac{135 H_0^2 \epsilon_0^2}{2 \theta_0^3 m^2_{\textbf{ph}}}-\frac{15 m^2_{\textbf{ph}}}{2 \theta_0^3 H_0^2}+\frac{18 H_0^2 \epsilon_0^2}{\theta_0^2 m^2_{\textbf{ph}!
 }}-\frac{18 H_0^2 \epsilon_0}{\theta_0^2 m^2_{\textbf{ph}}}-\frac{3 H_0^2}{2 \theta_0 m^2_{\textbf{ph}}}\bigg],\\ 
\Delta m^2&=\frac{\lambda_{\textbf{ph}} m_{\textbf{ph}}^2 }{16 \pi ^2}\bigg[-\frac{324 \epsilon_0^5}{\theta_0^4}-\frac{1296 \epsilon_0^4}{\theta_0^4}+\frac{3888 \epsilon_0^3}{\theta_0^4}-\frac{1944 \epsilon_0^2}{\theta_0^4}-\frac{90 \epsilon_0^3}{\theta_0^3}+\frac{1188 \epsilon_0^2}{\theta_0^3}\nonumber \\&-\frac{1008 \epsilon_0}{\theta_0^3}-\frac{18 \epsilon_0}{\theta_0^2}-\frac{81}{\theta_0^2}-\frac{72 m^4_{\textbf{ph}}}{\theta_0^4 H_0^4}-\frac{972 H_0^2 \epsilon_0^7}{\theta_0^4 m^2_{\textbf{ph}}}+\frac{972 H_0^2 \epsilon_0^6}{\theta_0^4 m^2_{\textbf{ph}}}+\frac{3888 H_0^2 \epsilon_0^5}{\theta_0^4 m^2_{\textbf{ph}}}\nonumber \\&-\frac{5832 H_0^2 \epsilon_0^4}{\theta_0^4 m^2_{\textbf{ph}}}+\frac{1944 H_0^2 \epsilon_0^3}{\theta_0^4 m^2_{\textbf{ph}}}-\frac{648 m^2_{\textbf{ph}} \epsilon_0^2}{\theta_0^4 H_0^2}+\frac{648 m^2_{\textbf{ph}} \epsilon_0}{\theta_0^4 H_0^2}+\frac{1782 H_0^2 \epsilon_0^4}{\theta_0^3 m^2_{\textbf{ph}}}-\frac{3564 H_0^2 \epsilon_0^3}{\theta_0^3 m^2_{\textbf{ph}}}\nonumber \\&+\frac{1512 H_0^2 \epsilon_0^2}{\theta_0^3 m^2_{\textbf{ph}}}+\frac{168 m^2_{\textbf{ph}}}{\theta_0^3 H_0^2}+\frac{27 H_0^2 \epsilon_0^3}{2 \theta_0^2 m^2_{\textbf{ph}}}-\frac{324 H_0^2 \epsilon_0^2}{\theta_0^2 m^2_{\textbf{ph}}}+\frac{243 H_0^2 \epsilon_0}{\theta_0^2 m^2_{\textbf{ph}}}+\frac{3 \theta_0 H_0^2 \epsilon_0}{4 m^2_{\textbf{ph}}}\nonumber \\&+\frac{9 H_0^2 \epsilon_0}{\theta_0 m^2_{\textbf{ph}}}-\frac{3 \theta_0 H_0^2}{2 m^2_{\textbf{ph}}}-\frac{9 H_0^2 \epsilon_0^3}{4 m^2_{\textbf{ph}}}+\frac{7 H_0^2 \epsilon_0^2}{m^2_{\textbf{ph}}}-\frac{7 H_0^2 \epsilon_0}{2 m^2_{\textbf{ph}}}-\frac{3 H_0^2}{m^2_{\textbf{ph}}}-\frac{\epsilon_0}{2}+1\bigg],\\ 
\Delta \eta &=-\frac{3 \lambda_{\textbf{ph}} ^2\varphi_0}{16 \pi ^2}\bigg[\frac{24\delta_0^2}{\theta_0^4}-\frac{20\delta_0}{\theta_0^3}+\frac{24 m_\textbf{ph}^4}{\theta_0^4H_0^4}-\frac{48\delta_0 m_{\textbf{ph}}^2}{\theta_0^4H_0^2}+\frac{20 m_{\textbf{ph}}^2}{\theta_0^3H_0^2}\bigg],\\ 
\Delta \lambda&=\frac{3 \lambda_{\textbf{ph}} ^2}{16 \pi ^2}\bigg[\frac{24\delta_0^2}{\theta_0^4}-\frac{24\delta_0}{\theta_0^3}+\frac{3}{\theta_0^2}+\frac{24 m_\textbf{ph}^4}{\theta_0^4H_0^4}-\frac{48\delta_0 m_{\textbf{ph}}^2}{\theta_0^4H_0^2}+\frac{24 m_{\textbf{ph}}^2}{\theta_0^3H_0^2}\bigg],\\
\Delta\xi&=\frac{\lambda_{\textbf{ph}} }{96 \pi ^2}\bigg[-\frac{972 \epsilon_0^6}{\theta_0^4}+\frac{972 \epsilon_0^5}{\theta_0^4}+\frac{270 \epsilon_0^3}{\theta_0^3}+\frac{27 \epsilon_0^2}{2 \theta_0^2}+\frac{54 \epsilon_0}{\theta_0^2}+\frac{3 \theta_0}{4}+\frac{9}{\theta_0}-\frac{324 m^2_{\textbf{ph}} \epsilon_0^4}{\theta_0^4 H_0^2}\nonumber \\&-\frac{90 m^2_{\textbf{ph}} \epsilon_0^2}{\theta_0^3 H_0^2}-\frac{18 m^2_{\textbf{ph}}}{\theta_0^2 H_0^2}-\frac{m^2_{\textbf{ph}}}{2 H_0^2}-\frac{9 \epsilon_0^2}{4}+\frac{5\epsilon_0}{2}+\frac{3}{2}\bigg],}
where 
\ee{\delta_0\equiv \f{m_{\textbf{ph}}^2+(\lambda_{\textbf{ph}}/2)\varphi_0^2}{H_0^2},\quad\qquad \theta_0\equiv \delta_0-3\epsilon_0+3\epsilon_0^2.}

\section{On the linear order $\delta$ and $\epsilon$ loop contributions}
\label{sec:line}
Let us define the linear part of the loop contribution, denoted by $\langle \hat{\phi}^2\rangle_{\rm lin}$ in (\ref{eq:eom1}), as a contribution that can be written as a Taylor series in $\delta$ and $\epsilon$ and is a simple polynomial in $H^2$: 
\ee{\langle \hat{\phi}^2\rangle_{\rm lin}=H^2\big[A+B\delta+C\epsilon+\mathcal{O}(\delta\epsilon,\epsilon^2,\delta^2)\big].}
Postulating that this is a gravity scalar, we can assume that the pure gravity part is a polynomial of the geometric tensors. The only term that can result in $\epsilon H^2$ is $R$, and hence from general covariance we can deduce that $A=-2C$. With the definitions (\ref{eq:defeps}) and (\ref{eq:defdel}) we then get by including only the the linear part of the loop $\langle \hat{\phi}^2\rangle_{\rm lin}$ in the field equation of motion (\ref{eq:eom1})
\ee{\bigg[-\Box +m'^2_0+\xi'_0R\bigg]\varphi+\f{\lambda'_0}{3!}\varphi^3=0,\label{eq:fe}}
where 
\ee{m'^2_0=m^2_0+\f{\lambda}{2}B m^2,\qquad \xi'_0=\xi_0+\lambda\frac{ 6 B \xi -C}{12},\qquad \lambda'_0=\lambda_0+\f{3}{2}B\lambda^2.\label{eq:refe}}
In conclusion, we see that the terms of linear order in $\delta$ and $\epsilon$ can be absorbed as shifts of the bare parameters and will leave no trace in the field equation when renormalization is performed.

Similarly, suppose first that the linear part of the energy density has the expansion
\ee{\langle \hat{T}^Q_{00}\rangle_{\rm lin}=H^4\big(D+E\delta+F\epsilon)+\mathcal{O}(\delta\epsilon,\epsilon^2,\delta^2).}
From the $\propto \varphi R$ term in (\ref{eq:fe}), by integrating with respect to $\varphi$ we can deduce the coefficient $\propto \varphi^2 G_{00}$ of the energy-density and get the relation
\ee{E=\f{-C + 6 B \xi}{2}.}
This allows us to write for the linear part of the full energy-density as
\ee{(T_{00})_{\rm lin}=(T_{00}^C)'-2\delta \alpha ' G_{00}-2 \beta'~^{(1)}H_{00}+\delta\epsilon_3'~^{(3)}H_{00}+\mathcal{O}(\delta\epsilon,\epsilon^2,\delta^2),}
where $(T_{00}^C)'$ is the classical energy-density with the redefined constants from (\ref{eq:refe}) and with 
\ee{\delta \alpha' =\f{-C + 6 B \xi}{12},\qquad \delta \beta' =\frac{3 \xi  (C - 6 B \xi)+F}{216},\qquad \delta\epsilon_3' =6 \xi  (-C + 6 B \xi)+D}
and where we have used the definition
\ee{~^{(3)}H_{00}\equiv \frac{1}{n-4} \left(\frac{4 }{3\left(2-3 n+n^2\right)} ~^{(1)}H_{00}-\frac{16}{3n(n-2)} ~^{(2)}H_{00}\right)=H^4+\mathcal{O}(n-4)\label{eq:3H00}.}
The expressions for $~^{(1)}H_{\mu\nu}$ and $~^{(2)}H_{\mu\nu}$ can be found in appendix \ref{sec:appA}. 
Similar arguments apply to the pressure components as well, and therefore the above results show that the terms of linear order in $\delta$ and $\epsilon$ can be absorbed in the renormalization and hence discarded in the calculation of $\langle \hat{T}^Q_{\mu\nu}\rangle$.

To summarize, the terms of linear order in $\delta$ and $\epsilon$ can be discarded in the calculation of the field equation and the energy-momentum tensor, since their contribution can be absorbed in the renormalization. Note however that only polynomial $H$-dependence is considered here and hence the conclusion is not valid for the terms involving logarithms $\log(H)$. 
Moreover, this discussion is only valid for the perturbative 1PI case. 

\section{Covariant conservation of the 2PI energy-momentum tensor}
\label{sec:appcov}
Consistency of our solutions requires that the energy-momentum tensor defined in (\ref{eq:T2PI}) satisfies the covariant conservation relation
\ee{\nabla^\mu T_{\mu\nu}=0,\label{eq:cov}}
For a diagonal energy-momentum tensor with no dependence on the spatial coordinates in a FRW space-time, the only non-trivial component in the equation (\ref{eq:cov}) is $\nu=0$. The derivative of the last line in (\ref{eq:T2PI}) gives negligible contributions of $\mathcal{O}(\epsilon)$ and higher, while the first two lines in (\ref{eq:T2PI}) give
\ea{\nabla^\mu T_{\mu 0}&=\partial_0\bigg[\f{1}{2}\dot{\varphi}^2+\f{\lambda}{12}\varphi^4+\f{M_{\rm 2PI}^2\left(M_{\rm 2PI}^2-2M^2\right)}{2\lambda}\bigg]+\dot{\varphi}\Box\varphi+\partial^\mu\varphi\nabla_\mu\partial_0\varphi\nonumber \\ &+\f{2\xi}{\lambda}\big(\nabla^\mu R_{\mu0}+R_{\mu0}\nabla^\mu-\Box\nabla_0+\nabla_0\Box\big)M_{\rm 2PI}^2.}
Using the standard commutator formula
\ee{-\Box\nabla_0+\nabla_0\Box=-R_{\mu0}\nabla^\mu,}
and the twice contracted Bianchi identity
\ee{\nabla^\mu R_{\mu0}=\nabla_0\f{R}{2},} we can write the above as
\ee{\nabla^\mu T_{\mu\nu}=-\dot{\varphi}\bigg[-\Box\varphi+M_{\rm 2PI}^2\varphi-\f{\lambda}{3}\varphi^3\bigg]+\f{\dot{M}_{\rm 2PI}^2}{\lambda}\big(M_{\rm 2PI}^2-M^2\big).\label{eq:tempcov}}
Since the last term in (\ref{eq:tempcov}) is purely quantum correction (it vanishes in the classical limit $M_{\rm 2PI}^2\rightarrow M^2$) and moreover $\dot{M}_{\rm 2PI}^2 \sim \mathcal{O}(\epsilon\,\delta_{\rm 2PI})$, this term is actually beyond our approximation for the loop (quantum) contributions. Therefore, requiring covariant conservation is consistent with the equation of motion (\ref{eq:feom2pi}).

\section{Renormalization of the gap equation}
\label{sec:GapA}

In order to derive the counter terms for the gap equation it is convenient to write the loop contribution $G(x,x)$ as
\ee{G(x,x)=\frac{-M^2_{\rm 2PI}+R/6}{8 \pi ^2 (4-n)}+ {\mathcal{F}},\label{eq:G}}
where $\mathcal{F}$ is finite, such that the equation (\ref{eff_mass_2PI}) can be written as
\ea{M^2_{\rm 2PI}&=m^2+\xi  +\frac{\lambda }{2}\varphi ^2+\lambda  \f{\mathcal{F}}{2}\nonumber \\&+\bigg\{\delta m^2_0+ \delta \xi_0  R+\frac{\delta \lambda_1 }{2}\varphi ^2-(\delta \lambda_1 +\lambda )\frac{M^2_{{\rm 2PI}}-{R}/{6} }{16\pi ^2 (4-n) }+\delta \lambda_1 \f{\mathcal{F}}{2}\bigg\}.\label{eq:M2PI}}
If we impose the condition that the expression in the curly brackets in (\ref{eq:M2PI}) vanishes, we can write the above as
\ee{M^2_{\rm 2PI} =m^2+\xi  +\frac{\lambda }{2}\varphi ^2+\lambda \f{\mathcal{F}}{2}\label{eq:gap_app},}
and
\ea{&\bigg[\delta m^2_0-(\lambda+\delta\lambda_1)\f{m^2}{16\pi^2(4-n)}\bigg]+R\bigg[\delta\xi_0 -(\lambda+ \delta\lambda_1)\f{(\xi-{1}/{6}) }{16\pi^2(4-n)}\bigg]\nonumber \\ &-\bigg(\f{\varphi^2}{2}+\f{\mathcal{F}}{2}\bigg)\bigg[(\lambda+\delta\lambda_1)\f{\lambda}{16\pi^2(4-n)}-\delta\lambda_1\bigg]=0.
\label{eq:count2pi_der}}
The former equation (\ref{eq:gap_app}) gives the finite gap equation (\ref{eq:gap}) for the self mass $M_{\rm 2PI}$, while the latter equation (\ref{eq:count2pi_der}) gives the following expressions for the counter terms in the limit $n\rightarrow 4$:
\ea{\label{eq:count2pi}\delta\lambda_1&={\f{\lambda^2}{16\pi^2(4-n)}}\bigg(1-\f{\lambda}{16\pi^2(4-n)}\bigg)^{-1},\nonumber \\ \delta m^2_0&={\f{m^2\lambda}{16\pi^2(4-n)}}\bigg(1-\f{\lambda}{16\pi^2(4-n)}\bigg)^{-1},\nonumber \\ \delta\xi_0&={\bigg(\f{\big(\xi-{1}/{6}\big)\lambda}{16\pi^2(4-n)}}-\frac{7}{72}\frac{\lambda }{16\pi ^2}\bigg)\bigg(1-\f{\lambda}{16\pi^2(4-n)}\bigg)^{-1}.} In the above we have included an $\mathcal{O}(n-4)$ contribution in $\delta\xi_0$ in order to obtain the same running for $\tilde{\xi}$ as for the other constants in (\ref{eq:run}).

\end{document}